\documentclass[journal]{IEEEtran}
\usepackage{cite}

\usepackage{caption}

\usepackage{amsmath}
\usepackage{amssymb}
\usepackage{booktabs} 

\usepackage{makecell}

\usepackage{algorithm}  
\usepackage{algpseudocode}

\ifCLASSINFOpdf
  \usepackage[pdftex]{graphicx}
\else
\fi
\begin{document}
%
\title{A Trust-Aware and Cost-Optimized Blockchain Oracle Selection Model with Deep Reinforcement Learning}
%
%
%


\author{Hengyang~Zhang,
        Shike~Li,~\IEEEmembership{Student~Member,~IEEE,}
        Hang~Bao, 
        Sixing~Wu,
        and~Jianbin~Li
\thanks{This work is supported by Research on Key Technologies to Support Network Operation of Distributed Energy Storage (5100-202199544A-0-5- ZN).}     
\thanks{Hengyang Zhang, Hang Bao, Sixing Wu and Jianbin Li are with the School of Control and Computer Engineering, North China Electric Power University, Beijing, China (e-mail: \{hyzhang, wusx, lijb87\}@ncepu.edu.cn, hangbao06@163.com).}
\thanks{Shike Li is with the Shanxi University(e-mail: lsk@sxu.edu.cn).}}

\maketitle

\begin{abstract}

The rapid development of blockchain technology has driven the widespread application of decentralized applications (DApps) across various fields. However, DApps cannot directly access external data and rely on oracles to interact with off-chain data. As a bridge between blockchain and external data sources, oracles pose potential risks of malicious behavior, which may inject incorrect or harmful data, leading to trust and security issues. Additionally, with the surge in data requests, the disparity in oracle trustworthiness and costs has increased, making the dynamic selection of the most suitable oracle for each request a critical challenge. To address these issues, this paper proposes a Trust-Aware and Cost-Optimized Blockchain Oracle Selection Model with Deep Reinforcement Learning (TCO-DRL). The model incorporates a comprehensive trust management mechanism to evaluate oracle reputation from multiple dimensions and employs an improved sliding time window to monitor reputation changes in real time, enhancing resistance to malicious attacks. Moreover, TCO-DRL uses deep reinforcement learning algorithms to dynamically adapt to fluctuations in oracle reputation, ensuring the selection of high-reputation oracles while optimizing node selection, thereby reducing costs without compromising data quality. We implemented and validated TCO-DRL on Ethereum. Experimental results show that, compared to existing methods, TCO-DRL reduces the allocation rate to malicious oracles by more than 39.10\% and saves over 12.00\% in costs. Furthermore, simulated experiments on various malicious attacks further validate the robustness and effectiveness of TCO-DRL.

\end{abstract}


%
\IEEEpeerreviewmaketitle

\section{Introduction}
%
%
%
%
\IEEEPARstart{B}{lockchain} technology, with its decentralized architecture, consensus mechanisms, and data encryption capabilities, has been widely applied in building distributed IoT systems, effectively preventing potential attacks and enhancing the overall security of IoT networks \cite{9358215}. Additionally, the introduction of smart contracts allows predefined rules to be automatically executed without third-party intervention, further promoting the deep integration of blockchain and IoT across various fields, including computation offloading \cite{9336659}, resource management \cite{8891803}, shipment management \cite{baygin2022blockchain}, inventory management \cite{ho2021blockchain}, and smart grids \cite{8234700}. However, this integration of blockchain and IoT also introduces a critical issue: due to the isolation of blockchain environments from the outside world, smart contracts cannot directly access off-chain data \cite{van2018blockchain}. Existing studies often assume that the "entities" responsible for blockchain and off-chain data interaction are fully trustworthy, but this assumption is unrealistic.

These entities that act as bridges between blockchain and external data are known as blockchain oracles \cite{al2019decentralized}. Oracles are responsible for collecting and verifying data from the outside and then transmitting it to the blockchain \cite{hassan2023trust}. However, oracles are not always trustworthy; they can be manipulated or act out of self-interest by providing incorrect or misleading data, causing smart contracts to execute based on false information, which can lead to potential trust and security issues. This problem not only reintroduces the centralization risk in blockchain but also poses the threat of malicious data being injected into the blockchain, undermining the trust foundation of the system \cite{gigli2023decentralized}.

Researchers have developed various trust models for blockchain oracles. Some of these rely on trusted execution environments (TEE), such as Intel software guard extensions (SGX) \cite{costan2016intel}, and use widely adopted TLS Notary to provide authenticity proofs, as in oracle systems like Towncrier and Provable \cite{zhang2016town}. However, hardware-based proofs do not fully guarantee tamper-proof data, and these methods are not suitable for scenarios requiring manual input or where digital data is inaccessible \cite{taghavi2023reinforcement}. On the other hand, some oracle communities (such as ChainLink, Witnet, and Augur) \cite{al2020trustworthy} evaluate the trustworthiness of oracles based on reputation scores, using reputation as the core of their trust model \cite{resnick2000reputation}. However, these communities often rely on simple historical behaviour records to calculate reputation scores, which cannot comprehensively or accurately reflect the actual behaviour of the oracles and need more robust and integrated trust management mechanisms for evaluation. 

At the same time, with the rapid development of decentralized applications (DApps) based on blockchain, the frequency of external data interactions is continuously increasing, and the number of data requests is sharply rising \cite{chainlink2024update}. Selecting the appropriate oracle nodes to handle a large number of data requests has become one of the major challenges oracle systems face. The work \cite{taghavi2023reinforcement} utilized a reinforcement learning algorithm known as the Bayesian multi-armed bandit to select oracle nodes. However, the Bayesian multi-armed bandit is typically simplify the state space to a probability distribution over different options, making them inadequate for dealing with complex and dynamic environments. In contrast, deep reinforcement learning (DRL) combines the perception capabilities of deep learning with the decision-making abilities of reinforcement learning. It can learn complex decision-making rules through continuous interaction with the environment, allowing the system to automatically learn from historical experiences, adapt to environmental changes, and gradually optimize strategies \cite{zhang2023cost}. Therefore, using DRL to design oracle node selection strategies presents a promising solution to address the challenges of complex and dynamic environments.

To address the aforementioned issues, we propose a trust-aware and cost-optimized blockchain oracle node selection model based on deep reinforcement learning (TCO-DRL). Specifically, we have designed a comprehensive trust management mechanism that evaluates the reputation of oracles from multiple dimensions (such as reliability, behavior, etc.). Additionally, TCO-DRL introduces an improved sliding time window mechanism to store and calculate the final reputation of oracles, enabling timely detection of behavioral changes and preventing malicious nodes from undermining the system's trust foundation over a prolonged period. Building on this trust management mechanism, we further propose a DRL-based oracle node selection algorithm that dynamically adapts to the changing state of data requests and oracle nodes. Since highly reputable oracles are often associated with higher costs, TCO-DRL also incorporates user budget constraints and adjusts strategies to select more cost-effective yet trustworthy oracle nodes. Through dynamic strategy adjustments, it ensures that DApps' operating costs are minimized without compromising data trustworthiness. Ultimately, TCO-DRL achieves an optimal balance between trust and cost. The main contributions of this paper are summarized as follows:
\begin{enumerate}
    \item We proposed TCO-DRL, a model that evaluates the reputation of oracles through a multi-dimensional trust management mechanism, enabling the system to more accurately assess the trustworthiness of nodes. By incorporating an improved sliding time window, TCO-DRL integrates both historical and short-term trust indicators, enhancing the system's ability to defend against malicious nodes while reducing computational overhead.
    \item We addressed the complex problem of oracle node allocation by introducing a deep reinforcement learning-based selection algorithm within TCO-DRL. TCO-DRL intelligently selects the most suitable nodes based on the complexity of data requests, the dynamic changes in oracle reputation, and the cost of each oracle. In addition, TCO-DRL considers the varying needs of data requesters, optimizing cost-efficiency while ensuring trust.
    \item We implemented and validated TCO-DRL in experiments, comparing it with existing solutions that involve noise and deteriorating conditions. The results demonstrate that TCO-DRL excels in selecting trustworthy and cost-efficient oracles. Furthermore, we deployed TCO-DRL on the Ethereum platform and simulated several typical attacks, further proving the model's effectiveness and robustness. The relevant code has been made publicly available at https://github.com/elpsylearning/TCO-DRL.
\end{enumerate}

The remainder of this paper is structured as follows: Section II provides an overview of related work. Section III presents the system architecture and sequence diagram. Section IV describes the components of the trust management scheme in detail. In Section V, we outline the process of selecting oracle node based on DRL. Section VI presents and analyzes the experimental results. Finally, Section VII concludes our work.


\section{RELATED WORK}
\subsection{Blockchain-based Trust Management for IoT}
IoT is characterized by high complexity and heterogeneity, constantly facing significant security and privacy threats \cite{ray2018survey}. Trust management technologies play a crucial role in establishing a reliable and effective security mechanism for IoT \cite{roman2011securing}. One potential approach to computing trust is reputation-based \cite{aaqib2023iot}, which aggregates collected trust information in a centralized or distributed manner to generate reputation scores \cite{pourghebleh2019comprehensive}. In IoT environments, employing reputation-based trust management techniques can enhance the trustworthiness between entities and detect abnormal activities \cite{alzaid2013reputation}. However, existing trust management schemes still face challenges such as single points of failure, data consistency maintenance, and the provision of a global trust view \cite{liu2023survey}. Consequently, more research is leveraging the security features of blockchain to address these issues.

In the fields of edge computing and fog computing, the work \cite{yang2023trusted} designed a decentralized trust management mechanism to evaluate the trustworthiness of edge nodes. It implemented a trusted MEC (T-MEC) framework using a directed acyclic graph structured blockchain (DAG blockchain) and distributed trust management (DTM) to resolve inconsistencies between on-chain and off-chain trust. The work \cite{kouicem2020decentralized} proposed a layered and scalable trust management scheme to assess the credibility of service providers, where trust information is computed and managed by fog nodes and propagated through a blockchain network to obtain a global view of trust data. In the supply chain sector, the work \cite{malik2019trustchain} introduced a three-layer trust management framework capable of dynamically scoring reputations to evaluate the trustworthiness of supply chain entities. By combining consortium blockchain with trust management technologies, it addresses trust challenges related to the data itself. Similarly, the work \cite{wu2022integrated} proposed a blockchain-integrated trust management model for assessing the trust of supply chain entities, providing an analysis based on a real-world scenario. In the vehicular networks domain, the work \cite{du2023blockchain} developed a trust management scheme based on Bayesian inference, integrating consortium blockchain to solve trust issues in inter-vehicle information sharing. Additionally, the work \cite{9336659} utilized three-valued subjective logic (3VSL) to comprehensively evaluate entities within vehicular networks, achieving more accurate reputations and combining blockchain for secure computation offloading.

\subsection{Deep Reinforcement Learning for IoT}
Due to the heterogeneity, complexity, and dynamic nature of IoT systems, problems framed as game theory or combinatorial optimization are often NP-hard \cite{uprety2020reinforcement}. Many studies have employed dynamic programming and heuristic algorithms to address these challenges, but these methods typically require numerous iterations to achieve satisfactory solutions \cite{huang2019deep}. In recent years, DRL has been widely applied in IoT systems because of its advantages in long-term performance optimization, real-time decision making, online learning without prior knowledge, and high scalability \cite{chen2021deep}.

In the field of industrial IoT, the work \cite{luo2023deep} developed a job scheduling model for hybrid flow shop scheduling on batch processing machines (HFSP-BPM) based on DRL, overcoming the challenge where a fixed search paradigm could not simultaneously meet the requirements for real-time processing and solution quality. In the work \cite{chen2020deep}, a DRL-based approach was proposed for dynamic resource management in industrial IoT, aiming to minimize the average task delay. This approach leverages DRL’s capability to handle high-dimensional state spaces, addressing the issues that arise from the dynamic and continuous nature of tasks in resource-constrained IIoT environments. In the domain of smart grids, the work \cite{chung2020distributed} introduced a distributed DRL-based method for intelligent load scheduling in residential smart grids, which not only protects household privacy but also reduces grid stress and household electricity costs. The work \cite{park2022multi} applied multi-agent reinforcement learning to solve the electric vehicle charging scheduling problem in smart grids, allowing for quick responses to user demands while reducing operational costs for charging station operators. In the field of intelligent transportation, the work \cite{koh2020real} proposed a DRL-based approach to develop vehicle navigation systems designed to meet the real-time demands of complex urban environments, thereby alleviating traffic congestion. The effectiveness of this solution was validated through simulations in nine real-world traffic scenarios. The work \cite{qian2019deep} utilized DRL’s adaptive learning capabilities to develop an intelligent electric vehicle charging navigation system, aiming to minimize total travel time and charging costs for electric vehicles.

The problem of selecting oracle nodes can be modeled as a decision optimization problem. Leveraging the advantages of DRL technology to select the most suitable oracle from many nodes for complex data requests is a viable solution. Based on our research, at the time of writing this paper, no existing work has applied DRL technology to the blockchain oracle domain. Our work is the first attempt in this area.
\subsection{Blockchain Oracles}
Blockchain oracles, which act as intermediaries between blockchain networks and external data, have received increasing attention in recent years. The work \cite{gigli2023decentralized} proposed a decentralized IoT global marketplace architecture based on blockchain technology and oracle networks, allowing users to purchase IoT device data through smart contracts. Additionally, to ensure data quality and credibility, a reputation algorithm was designed to evaluate data sources. The work \cite{kochovski2019trust} developed a blockchain-based trust management system for Edge-to-Cloud environments, using oracles to obtain external system monitoring data to support smart contract execution, thereby establishing trust between service providers and stakeholders. However, both studies assume that oracles will always provide honest and reliable services, which is overly optimistic.

Oracles can also behave maliciously, potentially causing significant damage to the blockchain network. Many studies have adopted reputation as the trust model for oracle systems. For instance, the work \cite{lu2021exploring} in the construction industry used smart construction objects (SCOs) as blockchain oracles to transmit real-world construction process data to the blockchain and utilized smart contracts to rate the reputation of oracles. There are also commercial oracle networks like Chainlink \cite{breidenbach2021chainlink} and Witnet \cite{de2017witnet}. However, these studies often evaluate oracle reputation based only on some past records. Chainlink derives reputation scores by recording on-chain performance history, such as average response latency. The work \cite{lu2021exploring}, similar to Witnet, assigns a high reputation score to oracles that return results matching those of the majority of oracles, while those returning different results receive lower scores. These approaches lacks more comprehensive trust management mechanisms for evaluating the reputation of oracle nodes.

Moreover, current research on how to select oracle nodes is still quite limited, with only the work \cite{taghavi2023reinforcement} employing intelligent mechanisms for the flexible selection of oracle nodes. The Table \ref{tab:1} summarizes the comparison between the BLOR proposed in the work \cite{taghavi2023reinforcement} and the TCO-DRL introduced in this paper. Overall, TCO-DRL offers the following advantages:

\begin{table*}[!t]
\centering
\caption{Comparison of the proposed TCO-DRL with BLOR.}
\label{tab:1}
\renewcommand\arraystretch{1.3}
\begin{tabular*}{0.8\linewidth}{lcc}
\toprule
\textbf{Features} & \textbf{BLOR\cite{taghavi2023reinforcement}} & \textbf{TCO-DRL} \\
\midrule 
 \textbf{Trust Model} & Reputation & Reputation \\
 \makecell[tl]{\\\textbf{Reputation Metric}} &  \makecell[tc]{\\Success and failure count} & \makecell[tc]{Reliability score \\ Behavior score \\token score}\\
 \textbf{Execution Architecture} & On-chain &  On-chain/Off-chain\\ 
 \textbf{Learning Method} & Online learning with prior knowledge  & Online learning without prior knowledge\\
 \textbf{Adaptive Learning Capability} & Weak & Strong\\
 \textbf{Scalability} & Low & High \\
 \textbf{Time Factor} &  NO & YES\\
 \textbf{Trust} & YES & YES \\
 \textbf{Cost} & YES & YES \\
 \textbf{Service Matching}& NO & YES\\
 \textbf{Attack Resistance} & NO & YES \\
\bottomrule
\end{tabular*}
\end{table*}

\subsubsection{A more comprehensive trust management scheme} TCO-DRL not only considers factors such as the oracle's success rate, interaction frequency, and average response time to compute a reliability score but also evaluates the oracle's behavior and staked tokens. Additionally, TCO-DRL incorporates the influence of a time factor, resulting in a more comprehensive trust management scheme.
\subsubsection{Independence from prior knowledge} TCO-DRL does not rely on any predefined models or prior distributions but instead learns optimal strategies directly through interaction with the environment and experience replay, enabling automatic adjustment and optimization of strategies. By using neural networks as function approximators, TCO-DRL can handle complex, high-dimensional state and action spaces, demonstrating strong learning capabilities and adaptability in dynamic and complex environments.
\subsubsection{Improved quality of service} TCO-DRL additionally takes into account the specific service requirements added by data requesters, striving to match these with oracles that offer the same services to enhance quality of service and user experience.
\subsubsection{Resistance to attacks} TCO-DRL is designed with attack resistance in mind, incorporating an improved sliding time window in its trust management scheme, which makes it difficult for malicious nodes to continuously engage in harmful actions. This enhances the system's security and robustness against malicious attack.


\section{The Proposed TCO-DRL Model}
In this section, we demonstrate the process by which TCO-DRL selects trustworthy and cost-effective oracle nodes to establish trust between data requesters and external data, presenting the architecture and sequence diagram of TCO-DRL for illustration.

\subsection{The Architecture of TCO-DRL}
\begin{figure}[!t]
\centering
\includegraphics[width=3in]{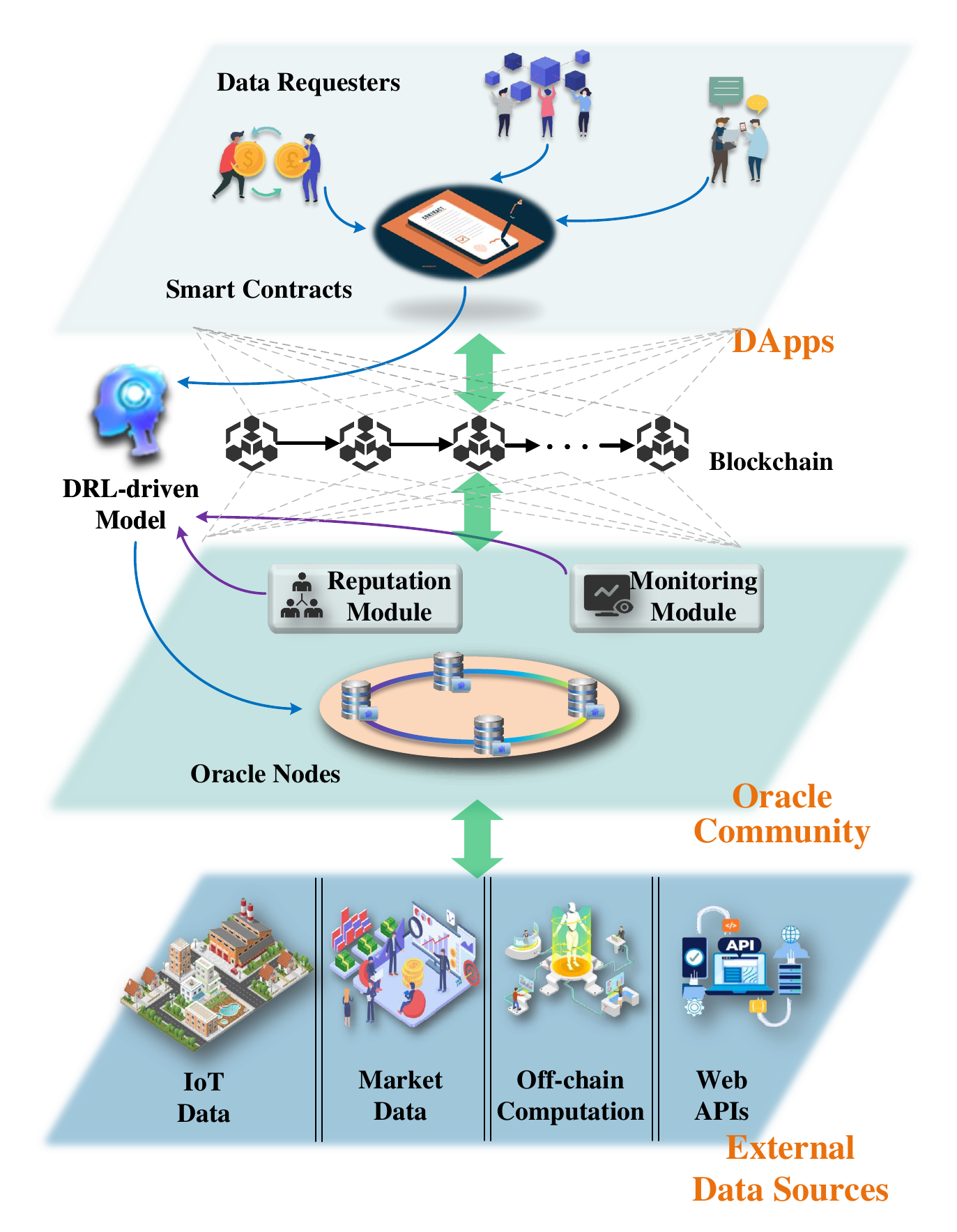}
\caption{The architecture of TCO-DRL.}
\label{fig_1}
\end{figure}

 As shown in Fig. \ref {fig_1}, the system architecture of TCO-DRL comprises three main entities: decentralized applications (DApps), oracle community, and external data sources. The descriptions of these entities are provided below.
\subsubsection{DApps}
DApps are applications that operate on blockchain technology, interacting with the blockchain network through smart contracts to deliver decentralized services and functionalities. These applications initiate data requests, requesting services from the oracle community to obtain off-chain data. The acquired data is subsequently used in the execution of smart contracts, enabling automated contract operations and decision-making.
\subsubsection{Oracle Community}
The oracle community acts as a bridge between the blockchain and the external world, comprising a network of oracle nodes responsible for retrieving data from external sources and transmitting it to the blockchain. Additionally, the oracle community includes monitoring and reputation modules, which gather information about the on-chain activities of oracles and employ reputation evaluation and verification mechanisms to ensure the accuracy and reliability of the data.
\subsubsection{External Data Sources}
External data sources refer to various off-chain data providers that supply input to oracle nodes. These sources may include IoT device data, financial market prices, web APIs, and complex off-chain computational tasks. Oracle nodes extract relevant information from these sources and transmit it to the blockchain, thereby providing the necessary external data support for the operation of DApps.

\begin{figure*}[!t]
\centering
\includegraphics[width=0.9\textwidth]{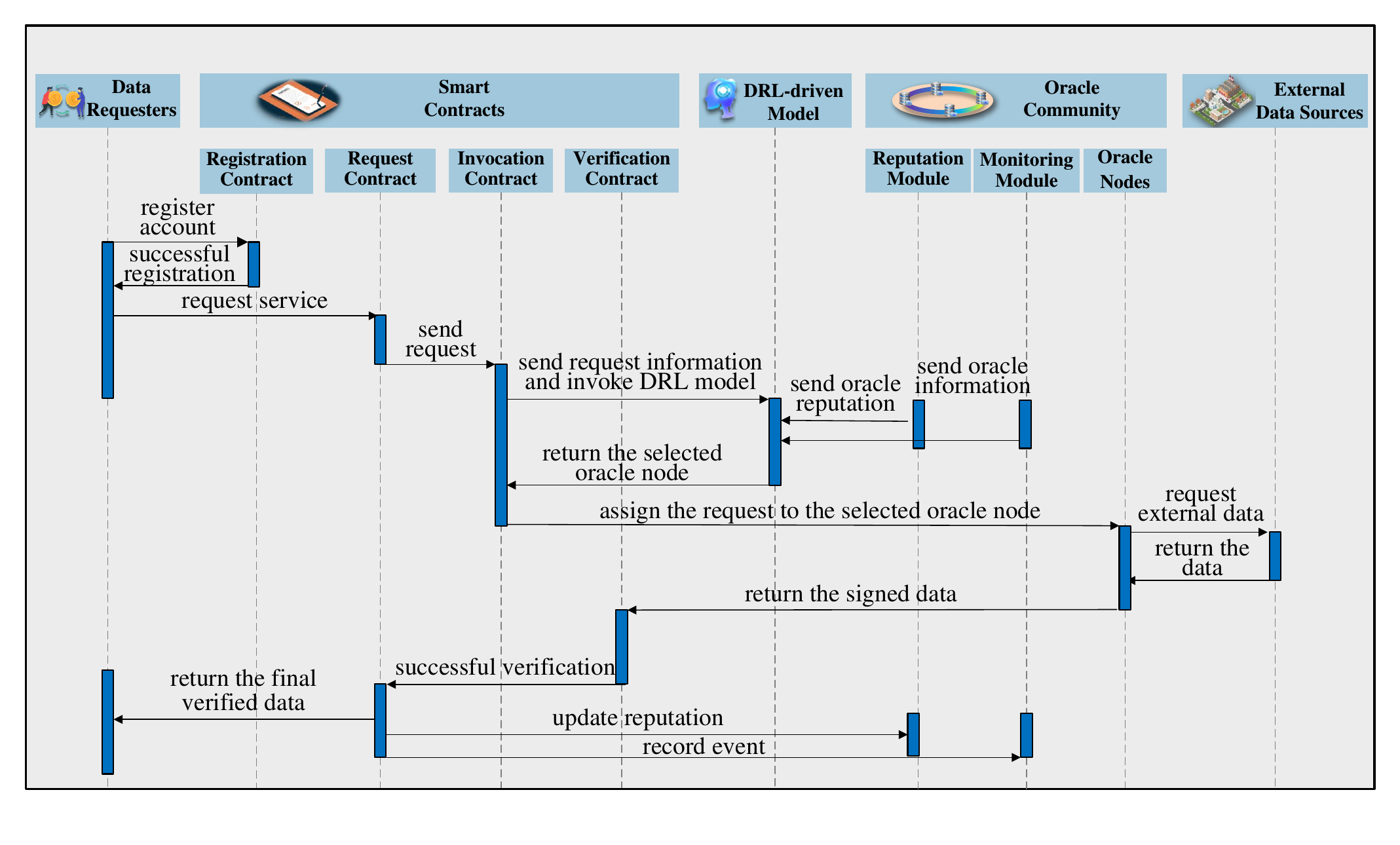}
\caption{The sequence diagram of TCO-DRL.}
\label{fig_2}
\end{figure*}

The major components of the system are detailed below.
\paragraph{Data Requesters}
Data requesters are entities that initiate data requests. They request services through smart contracts. Additionally, they may specify particular requirements for the data.
\paragraph{Smart Contracts}
The system includes four contracts: registration contract, request contract, invocation contract, and verification contract. The registration contract handles user account registration. The request contract defines data requests and attaches parameters such as data requirements. The invocation contract is responsible for loading and executing the DRL model to select oracle nodes. The verification contract is tasked with verifying data signatures and the validity of the data.
\paragraph{DRL-driven Model}
DRL-driven model is employed to optimize the selection of nodes within the oracle community. The Deep Q-Network (DQN) algorithm is used to dynamically choose the most appropriate nodes for providing services, based on on-chain data requests and the status information of the oracles. 
\paragraph{Reputation Module}
The reputation module is a critical component of the oracle community, responsible for evaluating and recording the reputation scores of each oracle node. In practice, the oracle community typically manages reputation scores through one or more smart contracts. Here, we refer to this component as the 'reputation module' for the sake of conceptual clarity.
\paragraph{Monitoring Module}
The monitoring module is responsible for overseeing the behavior and performance of oracle nodes. It continuously tracks the activities of these nodes, including their response times, the accuracy of data transmission, their online status, and so on.
\paragraph{Oracle Nodes}
Oracle nodes are the fundamental execution units within the oracle community, responsible for retrieving data from external sources and transmitting it to smart contracts on the blockchain. Each oracle node can operate independently and may be managed by different entities, such as individuals or organizations.
\subsection{The Sequence Diagram of TCO-DRL}
To further elaborate on the operation of TCO-DRL, we provide a detailed explanation of its specific process using the sequence diagram shown in Fig. \ref{fig_2}.

\begin{enumerate}
    \item Data requesters complete the registration process through the registration contract. 
    \item Data requesters submit service requests for external data through the request contract.
    \item The service request triggers the invocation contract. The invocation contract transmits the service request information to the DRL-driven Model and invokes the model for processing.
    \item Based on the service request information sent by the invocation contract, along with the latest reputation and relevant status information of oracle node obtained from the reputation module and monitoring module, the DRL-driven model selects trustworthy and cost-effective oracle node to handle the data request and returns the selection result to the invocation contract.
    \item The invocation contract sends the data request to the selected oracle node based on the selection result.
    \item The selected oracle node interact with external data sources to obtain the requested data.
    \item The external data sources transmit the data to the oracle node, which then sign the data and prepare it for submission to the blockchain.
    \item The oracle node return the signed data to the verification contract. The verification contract validates the data, and upon confirmation of its validity, notifies the request contract of the successful verification.
    \item The request contract delivers the final verified data back to the data requester, completing the entire data request process. The reputation module updates the reputation score of the oracle node based on its performance in this service. The monitoring module logs all operations related to this event and continues to monitor the subsequent performance of the oracle nodes.
\end{enumerate}

\section{The Proposed Trust Management Mechanism in TCO-DRL}
In this section, we provide a detailed explanation of the trust management mechanism designed for TCO-DRL to evaluate the trustworthiness of oracle nodes. Initially, the base reputation score for each oracle node is established by considering three dimensions: reliability score, behavior score, and token score. The integration of these three factors allows for the effective identification of nodes that are not only reliable and compliant but also demonstrate a strong sense of responsibility toward the network, thereby contributing to the security and stability of the oracle system. Additionally, the final reputation score of an oracle node is derived by incorporating a time factor into the base reputation score, ensuring that the system accurately reflects the node's current performance in real-time. This approach effectively prevents trust monopolies and reduces the risk of misjudgment due to historical data, thereby enhancing the robustness and security of the entire mechanism. The notations and relevant definitions can be found in Table \ref{tab:2}.

\subsection{Reliability Score}
The reliability score directly reflects the performance of an oracle node in executing its core function—providing data. This score aggregates three key metrics: relative response frequency, success rate, and average response time. Combining these three metrics helps to mitigate uncertainties caused by deviations in a single metric, offering a more comprehensive insight into the node’s performance. Ultimately, this approach aids in distinguishing nodes capable of consistently delivering high-quality data from those that may occasionally provide incorrect or delayed information. The detailed calculation process for each metric is outlined below.

\subsubsection{Relative Response Frequency}
Relative response frequency reflects the activity level and responsiveness of an oracle within the network. A higher relative response frequency indicates that the oracle is highly available and frequently participates in processing data requests. This metric is determined by calculating the ratio of the number of times the oracle actually responds to data requests within a specific time period to the average response count of all oracles:
\begin{equation}
ORF_j(\tau_k) = N_j(\tau_k) \times \frac{M}{\sum_{m=1}^{M} N_m(\tau_k)}
\end{equation}
Where $ ORF_j(\tau_k) $ represents the relative response frequency of the oracle $j$ during the time period $\tau_k$. $ N_j(\tau_k) $ denotes the number of data requests the oracle $j$ responded to within the time period $\tau_k$. $M$ indicates the total number of oracles within the oracle community.

\subsubsection{Success Rate}
Success rate is a key metric for evaluating the accuracy and effectiveness of the data provided by an oracle. Oracles with a high success rate typically demonstrate strong data processing capabilities, consistently delivering accurate and valid data in most cases. It is calculated as the ratio of the number of data requests successfully processed by the oracle to the total number of requests it has responded to:
\begin{equation}
OSR_j(\tau_k) = \frac{\sum_{i=1}^{N} y_{i,j}(\tau_k) \cdot Osucc_{i,j}(\tau_k)}{N_j(\tau_k)}
\end{equation}
Where $ y_{i,j}(\tau_k) $ is an indicator variable, taking the value of 1 if request $i$ is assigned to oracle $j$ during time period $\tau_k$, and 0 otherwise. $Osucc_{i,j}(\tau_k)$ indicates whether request $i$ was successfully completed by oracle $j$ during time period $\tau_k$, with $Osucc_{i,j}(\tau_k)=1$ for success and $Osucc_{i,j}(\tau_k)=0$ for failure. $N_j(\tau_k)$ denotes the total number of requests processed by oracle $j$ during time period $\tau_k$.The success of processing a data request is determined using the following equation:
\begin{equation}
Osucc_{ij}(\tau_k) = \\
\begin{aligned}
\begin{cases} 
1, & \text{if } ORT_{ij}(\tau_k) \leq DDL_i \text{ and } \\ 
 & Verification_{ij} = \text{TRUE} \\
0, & \text{else}
\end{cases}
\end{aligned}
\end{equation}
Where $ORT_{ij}(\tau_k)$ represents the response time of oracle $j$ in processing data request $i$ during time period $\tau_k$. $DDL_i$ denotes the deadline for data request $i$. $Verification_{ij}$ indicates whether the data request $i$ processed by oracle $j$ has passed verification.

\subsubsection{Average Response Time}
Average response time measures the efficiency of an oracle in processing and providing data. A shorter response time indicates a faster processing speed, enabling the oracle to provide data more promptly. Selecting oracles based on average response time helps improve the overall system's responsiveness and the real-time availability of data. It is calculated using the following equation:
\begin{equation}
ORT_j(\tau_k) = \frac{\sum_{i=1}^{N} y_{i,j}(\tau_k) \cdot (TS_{finish_{ij}} - TS_{start_{ij}})}{N_j(\tau_k)}
\end{equation}
Where $TS_{finish_{ij}}$ represents the timestamp when oracle $j$ completes request $i$, and $TS_{start_{ij}}$ denotes the timestamp when request $i$ begins processing by oracle $j$. $N_j(\tau_k)$ denotes the total number of requests processed by oracle $j$ during time period $\tau_k$.

Aggregating the above three metrics results in the reliability score of oracle $j$ for time period $\tau_k$:
\begin{equation}
\begin{aligned}
\text{reliability score}_j(\tau_k) = & \omega \cdot ORF_j(\tau_k) + \varphi \cdot OSR_j(\tau_k) \\
&+ \psi \cdot \left(\frac{DDL}{ORT_j(\tau_k)}\right)
\end{aligned}
\end{equation}
Where $\omega$, $\varphi$ and $\psi$ represent the weights assigned to the relative response frequency, success rate, and average response time, respectively, with $\omega + \varphi + \psi = 1$. $DDL$ denotes the deadline for the data request, which is used to assess the efficiency level of the oracle by comparing it to the oracle’s average response time.

\begin{table}[!t]
\centering
\caption{Table of abbreviated parameters}
\label{tab:2}
\begin{tabular}{ll}
\toprule
\textbf{Symbol} & \textbf{Definition} \\
\midrule 
$ N_j(\tau_k) $ & \makecell[tl]{The number of data requests the oracle $j$ \\  responded to during time period $\tau_k$}\\
$ M $   & The total number of oracles in the oracle community \\
$ y_{i,j}(\tau_k) $ &  \makecell[tl]{Denotes whether request $i$ is assigned to oracle $j$ \\ during time period $\tau_k$} \\
$Osucc_{i,j}(\tau_k)$  &  \makecell[tl]{Denotes whether request $i$ is successfully \\ completed by oracle $j$ during time period $\tau_k$}\\
$ORT_{ij}(\tau_k)$ &  \makecell[tl]{The response time of oracle $j$ in processing \\ data request $i$ during time period $\tau_k$}\\ 
$DDL_i$ & The deadline of data request $i$\\
$Verification_{ij}$ &  \makecell[tl]{Denotes whether the data request $i$ processed \\ by oracle $j$ has passed verification}\\
$TS_{start_{ij}}$ & \makecell[tl]{The timestamp when request $i$ begins processing\\ by oracle $j$}\\
$ TS_{finish_{ij}}$ & The timestamp when oracle $j$ completes request $i$\\ 
$ ORF_j(\tau_k) $ & \makecell[tl]{The relative response frequency of the oracle $j$ \\ during time period $\tau_k$}\\
$OSR_j(\tau_k)$ & The success rate of the oracle $j$ during time period $\tau_k$\\
$ORT_j(\tau_k)$ & \makecell[tl]{The average response time of the oracle $j$\\ during time period $\tau_k$}\\
$SF$ & The harm score of safe behavior\\
$MH$ & The harm score of minor harm behavior\\ 
$MoH$ & The harm score of moderate harm behavior\\ 
$SH$  & The harm score of severe harm behavior\\ 
$n_{SF,j}(\tau_k)$ & \makecell[tl]{The number of times oracle $j$ exhibited \\ safe behavior during time period $\tau_k$}\\
$n_{MH,j}(\tau_k)$ & \makecell[tl]{The number of times oracle $j$ exhibited \\ minor harm behavior during time period $\tau_k$}\\
$n_{MoH,j}(\tau_k)$ & \makecell[tl]{The number of times oracle $j$ exhibited\\ moderate harm behavior during time period $\tau_k$}\\
$n_{SH,j}(\tau_k)$  & \makecell[tl]{The number of times oracle $j$ exhibited \\ severe harm behavior during time period $\tau_k$}\\
$\text{Token}_j(\tau_k)$ & \makecell[tl]{The number of tokens staked by oracle $j$ \\ during time period $\tau_k$}\\
$\text{Oreputation}_j(\tau_k)$ & \makecell[tl]{The final reputation score of oracle $j$ \\ during time period $\tau_k$}\\
\bottomrule
\end{tabular}
\end{table}

\subsection{Behavior Score}
The behavior score primarily evaluates the overall conduct of an oracle within the network. Oracles that exhibit good behavior are generally considered more trustworthy and contribute positively to the long-term health of the network. This score is crucial for promptly identifying oracles engaging in non-compliant or malicious activities, thereby enhancing the system's security and stability. In the design of the trust management mechanism presented in this paper, preventing deliberate malicious actions by oracles is a key objective. Therefore, if any malicious behavior is detected, the oracle's reputation score will be significantly reduced, dropping below the trust threshold, which will result in the oracle being disallowed from providing further services. Specifically, we categorize oracle behavior into four levels: Safe, Minor Harm, Moderate Harm, and Severe Harm, drawing on risk assessment models. Detailed descriptions of these levels are provided below.

\begin{figure}[!t]
\centering
\includegraphics[width=2.5in]{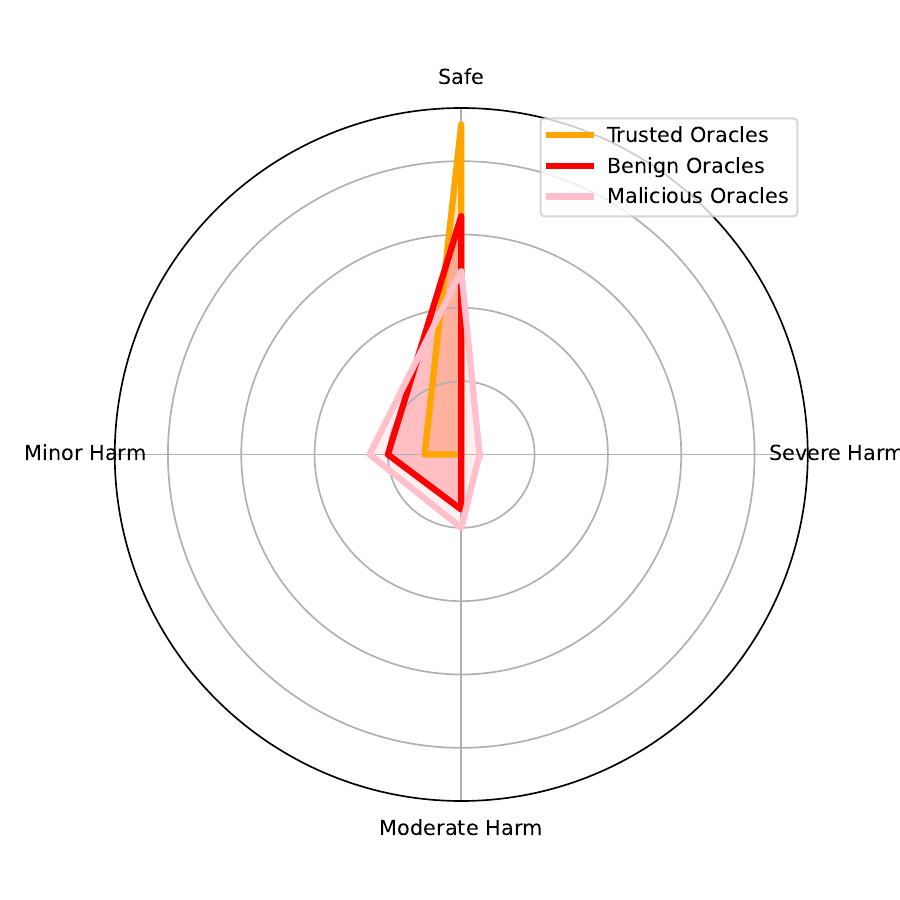}
\caption{Behavioral probabilities of various oracles.}
\label{fig_3}
\end{figure}

\subsubsection{Safe}
Safe indicates that the oracle has not engaged in any behavior during its operation that could negatively impact the system or other participants. The oracle has performed well during this period, providing accurate and timely data, actively participating in the consensus process, and adhering to all established rules and protocols.
\subsubsection{Minor Harm}
Minor harm indicates that the oracle may have committed minor violations or errors under certain circumstances, which do not pose a significant threat to the system’s overall security and stability but may affect certain functionalities or service quality. Examples include data delays or localized errors caused by software defects or minor performance issues.
\subsubsection{Moderate Harm}
Moderate harm suggests that the oracle’s behavior carries certain risks, potentially affecting some system functions or causing direct economic losses or trust issues for a subset of users. Examples include frequently providing inaccurate data, occasional violations of protocols, or failing to complete critical tasks within the designated time frame.
\subsubsection{Severe Harm}
Severe harm indicates that the oracle’s actions pose a threat to the system, endangering its stability and security, potentially leading to data loss, system crashes, serious security vulnerabilities, or irreversible damage to other nodes. Examples include the provision of widespread false data to manipulate markets, executing malicious code, intentionally violating protocols, or repeated severe violations. 

Ultimately, the behavior score of oracle $j$ during time period $\tau_k$ is calculated as follows:
\begin{equation}
\begin{aligned}
\text{behavior score}_j(\tau_k) = & SF \cdot n_{SF,j}(\tau_k) + MH \cdot n_{MH,j}(\tau_k) \\
& + MoH \cdot n_{MoH,j}(\tau_k) \\
& + SH \cdot n_{SH,j}(\tau_k)
\end{aligned}
\end{equation}
Where $SF$, $MH$, $MoH$, and $SH$ represent the harm scores for the four behavior categories: safe, minor harm, moderate harm, and severe harm, respectively. $n_{SF,j}(\tau_k)$, $n_{MH,j}(\tau_k)$, $n_{MoH,j}(\tau_k)$, and $n_{SH,j}(\tau_k)$ denote the number of times oracle $j$ exhibited each of these behaviors during time period $\tau_k$. Fig. \ref{fig_3} illustrates the probabilities of different behaviors for the three types of oracles simulated in our experiment: trustworthy oracles, benign oracles, and malicious oracles.

\subsection{Token Score}
The core function of the token score is to establish an economic constraint mechanism for oracles. By requiring oracles to stake a certain amount of tokens during their participation, the cost of not following the rules or providing incorrect data is effectively increased. A higher stake means that the oracle’s actions in the system carry greater economic risk, providing a stronger incentive for the oracle to maintain good behavior and deliver accurate data, in order to protect its staked tokens from being forfeited, thereby reducing the likelihood of malicious actions. Additionally, calculating and evaluating the token score is relatively straightforward, transparent, and can be done by simply referring to the oracle's staking records. This calculation imposes no significant burden on system performance while enhancing the system's credibility. The token score is calculated as the ratio of the number of tokens staked by the oracle to the average number of tokens staked by all oracles in the community:
\begin{equation}
\text{token score}_j(\tau_k) = \text{Token}_j(\tau_k) \times \frac{M}{\sum_{m=1}^{M} \text{Token}_m(\tau_k)}
\end{equation}
Where $M$ represents the number of oracles in the community, and $\text{Token}_j(\tau_k)$ denotes the number of tokens staked by oracle $j$ during time period $\tau_k$.

By aggregating the reliability score, behavior score, and token score, the base reputation value of oracle $j$ for time period $\tau_k$ is determined as follows:
\begin{equation}
\begin{aligned}
\text{base reputation}_j(\tau_k) = & \xi \cdot \text{reliability score}_j(\tau_k) \\
& - \zeta \cdot \text{behavior score}_j(\tau_k) \\
& + \delta \cdot \text{token score}_j(\tau_k)
\end{aligned}
\end{equation}
Where $\xi$, $\zeta$, and $\delta$ are the weighting coefficients for the reliability score, behavior score, and token score, respectively, with $\xi + \zeta + \delta = 1$.

\subsection{Time Factor}
The time factor is an element used to adjust and manage the impact of time on reputation scores. By incorporating the time factor into the trust management mechanism, the system can more rapidly respond to changes in oracle behavior and promptly adjust trust relationships. This paper adopts the function $\tanh\left(\frac{1}{x}\right)$, as utilized in the work \cite{li2023td}, as the time factor for TCO-DRL. The curve of this function is well-suited for influencing reputation, allowing the system to effectively capture recent behaviors and prevent short-term misconduct from being overlooked, while also balancing historical performance to avoid letting past behavior completely dominate the current reputation score. Additionally, we introduced a weighting coefficient $\chi$ to adjust the time factor, making its variation curve better suited to the studied scenario. The final reputation score of oracle $j$ for time period $\tau_k$ is calculated as follows:
\begin{equation}
\text{Oreputation}_j(\tau_k) = \sum_{\tau_1}^{\tau_k} \tanh\left(\frac{\chi}{\tau_k}\right) \cdot \text{base reputation}_j(\tau_k)
\end{equation}

\begin{table}[!t]
\centering
\caption{The final reputation of 5 oracles after 100 time periods with different lengths of sliding time window.}
\label{tab:3}
\begin{tabular}{cccccc}
\toprule
\textbf{Length} & \textbf{Oracle 0} & \textbf{Oracle 1} & \textbf{Oracle 2} & \textbf{Oracle 3} & \textbf{Oracle 4} \\
\midrule 
 1 & -0.06935 & -0.29814 & 0.20315 & 0.17370 & 0.51310 \\
 2 & -7.32857 &  0.07426 & -2.02773 & 0.33217 & 0.69670 \\
 3 & -24.38311 & 0.33590 & -0.44883 & 0.78209 & 0.87278 \\
 4 & -2.35772 & 0.46118 & -0.20747 & 1.12951 & 1.28135 \\
 \textbf{5}  & \textbf{-10.32644}  & \textbf{0.98567}  & \textbf{-0.71223}  & \textbf{1.33996}  & \textbf{2.24718} \\
 6 & -11.09053 & 1.45541 & 1.07729 & 3.58305 & 5.02434 \\ 
 7 & -67.42975 & 5.51652 & -1.39116 & 9.83602 & 12.29896 \\
 8 & -291.00294 & 14.09663 & -9.62620 & 25.88607 & 34.86167 \\
 9 & -839.47177 & 23.50501 & -49.98351 & 62.04435 & 78.97845 \\
 10 & -1802.25521 & 53.32101 & -99.06507 & 101.07012 & 144.07206 \\
\bottomrule
\end{tabular}
\end{table}

It is important to note that we use a sliding time window approach to store and calculate reputation scores. However, after testing, we found that the standard sliding time window approach is not well-suited for blockchain systems, leading us to make specific modifications.
\begin{itemize}
    \item Problem: Typically, a sliding time window records only the reputation of the oracle as assessed within that specific time period, with each window being independent and unconnected to others. The issue with this design is that if the sliding time window is short, the impact of malicious behavior on the oracle's reputation quickly dissipates, allowing a malicious oracle to easily regain the system's trust. On the other hand, extending the length of the sliding time window adds additional computational burden to the already resource-constrained blockchain network, thereby increasing system latency.
    \item Solution: Based on the considerations above, we record not just the independent reputation score within each time window, but rather a composite reputation score that incorporates the time factor and combines values from other time windows. This approach allows for a shorter sliding time window, reducing the need for maintaining extensive historical records and thus decreasing the computational burden on the blockchain system. At the same time, it extends the impact of malicious behavior on an oracle's reputation, ensuring that an oracle that has committed wrongdoing will require a significantly longer period before it can rejoin the community and continue providing services. The experimental results in Section VI demonstrate the effectiveness and reliability of our improvements.
\end{itemize}
We determined the appropriate length of the sliding time window used in TCO-DRL through experimentation. As shown in the Table \ref{tab:3}, we tested the final reputation values of 5 oracles after 100 time periods with different lengths of sliding time windows. The results indicate that if the sliding time window is too short, it can still differentiate between malicious and trusted oracles, but the distinction between different trusted oracles becomes less clear. Additionally, if the sliding time window is too long, it results in excessive accumulation of reputation scores, leading to exponential growth. Ultimately, we set the length of the sliding time window to 5.

\section{Trust and Cost-Aware Oracle Node Selection Algorithm in TCO-DRL}
In this section, we provide a detailed explanation of the specific algorithmic process implemented in TCO-DRL for selecting trustworthy and cost-effective oracles. We begin by introducing the underlying algorithm employed in our model—the Deep Q-Network (DQN). Next, we describe the DRL modeling of the problem. Finally, we present the specific process for dynamically selecting oracles based on DQN.
\subsection{Deep Q-Network}
Deep Q-Network (DQN) is a deep reinforcement learning algorithm that builds upon Q-Learning. Q-Learning \cite{watkins1992q} works by iteratively updating a function known as the Q-value, $Q(s, a)$, to learn the long-term rewards associated with taking a specific action $a$ in a given state $s$. However, traditional Q-Learning relies on a Q-table to store the Q-values for all state-action pairs, which becomes impractical in high-dimensional state spaces, as the size of the Q-table increases exponentially with the complexity of the state and action spaces. To address these limitations, DQN introduces several key improvements \cite{mnih2015human}, including the following:
\begin{itemize}
    \item Approximating the Q-Value Function Using DNNs: The core idea of DQN is to enhance Q-Learning by introducing Deep Neural Networks (DNNs) to approximate the Q-value function, which predicts the Q-values for state-action pairs. This approach addresses the issue of traditional Q-Learning being ineffective in high-dimensional state spaces.
    \item Experience Replay: DQN stores the state transitions experienced by the agent in a replay memory, from which it randomly samples small batches of data for training. This process breaks the correlation between data points, stabilizes the training process, and enhances data efficiency.
    \item Target Network: DQN employs two neural networks to stabilize the training process. One is the evaluation network, which is used for actual Q-value updates, and the other is the target network, which is used to calculate the target values for Q-value updates. The parameters of the target network are periodically copied from the evaluation network to reduce training instability.
\end{itemize}

\subsection{DRL Model Formulation}
Before introducing the specifics of DRL modeling, we first define the characteristics of data requests and oracles. Suppose there are $N$ data requests, denoted as $Request_1$, $Request_2$, ..., $Request_N$. For a data request $Request_i$, $ i \in \{{1, 2,..., N}\} $, it can be modeled by the following characteristics:
\begin{equation}
\begin{aligned}
\text{Request}_i = \{ & \text{RID}_i, \text{RarrivalTS}_{i,j}, \text{DDL}_i, \\
                    & \text{Rcomplexity}_i, \text{Rservice}_i \}
\end{aligned}
\end{equation}
Where $\text{RID}_i$ represents the unique ID that identifies data request $Request_i$. $\text{RarrivalTS}_{i,j}$ denotes the timestamp when data request $Request_i$ reaches oracle $j$. $\text{DDL}_i$ indicates the allowed deadline for $Request_i$, and $\text{Rcomplexity}_i$ represents the complexity of $Request_i$. $\text{Rservice}_i$ refers to some additional service requirements for $Request_i$, such as data update frequency and data accuracy.

Similarly, suppose there are $M$ oracles in the oracle community, denoted as $Oracle_1$, $Oracle_2$, ..., $Oracle_M$. The $Oracle_j$, $ j \in \{{1, 2,..., M}\} $ can be expressed as the following characteristics:

\begin{equation}
\begin{aligned}
\text{Oracle}_j = \{ & \text{OID}_j, \text{Oreputation}_j, \text{Ocost}_j, \\
                    & \text{Operformance}_j, \text{Oservice}_j \}
\end{aligned}
\end{equation}
Where $\text{OID}_j$ represents the unique ID that identifies $Oracle_j$, $\text{Oreputation}_j$ denotes the reputation score of $Oracle_j$. $\text{Ocost}_j$ indicates the cost per request for $Oracle_j$, $\text{Operformance}_j$ represents the processing capability of $Oracle_j$ for handling data requests, and $\text{Oservice}_j$ refers to the services that $Oracle_j$ can provide.

\subsubsection{State Space}
The state space represents the complete set of all possible conditions that an agent can perceive in its environment. Each state captures a snapshot of the environment at a specific moment, encompassing all the information necessary for the agent to make decisions. In TCO-DRL, the state space consists of the states of both data requests and oracles, denoted as $S = [S_{Request} \cup S_{Oracle}] $.
\subsubsection{Action Space}
The action space refers to the set of all possible actions an agent can choose from in any given state. Since the action in TCO-DRL involves selecting an oracle from the oracle community to provide service, the action space is defined as the set of oracles, i,e., $A = [OID_1, OID_2, \ldots, OID_M]$.
\subsubsection{Reward Function}
The reward function is the feedback signal received by the agent after interacting with the environment, indicating the quality of an action taken and guiding the agent in learning the optimal policy. The reward function designed in TCO-DRL is as follows:
\begin{equation}
\begin{aligned}
R &= \left(1 + \vartheta \cdot e^{\lambda - cost_j}\right) 
\left(\frac{exeT_{ij}}{responseT_{ij}}\right) \\
&\quad + Oreputation_j - \mu \cdot penalty_{ij}
\end{aligned}
\end{equation}

Where the conditions for the value of variable $penalty_{ij}$ are as follows:
\begin{equation}
penalty_{ij} = 
\begin{cases} 
0, & \text{if } Rservice_i = Oservice_j \\ 
1, & \text{else} 
\end{cases}
\end{equation}

In summary, the reward is directly proportional to the reputation $Oreputation_j$ of oracle $j$ and inversely proportional to the cost $cost_j$ of oracle $j$, guiding the agent to select oracles with high reputation and low cost to handle data requests. The reward function also encourages the service $Oservice_j$ provided by oracle $j$ to align closely with the additional service requirement $ Rservice_i $ of data request $i$, with mismatches resulting in penalty $penalty_{ij}$. Furthermore, to reduce the waiting time for data requests in queue and improve service quality, we also incorporate the proportion of the request execution time in the total response time into the reward function. The $\vartheta$, $\lambda$, and $\mu$ are the weighting coefficients for adjusting the influence of $\frac{exeT_{ij}}{responseT_{ij}}$, $cost_j$, and $penalty_{ij}$, respectively.

\subsection{Dynamic Oracle Node Selection Based on DQN}
Next, we provide a detailed explanation of the specific process for dynamically selecting oracles based on the DQN algorithm, including both the training phase and the execution phase. Fig. \ref{fig_4} visually illustrates the selection process.
\subsubsection{Initialize the DQN Model} We begin by initializing an evaluation network to approximate the Q-value function $Q(s, a; \theta)$, where $s$ represents the state, $a$ represents the action, and $\theta$ denotes the network parameters. We set the parameters for the evaluation network, including the exploration rate $\epsilon$, the learning rate $ \alpha $, and the discount factor $ \gamma $. Next, we initialize the target network, which has the same structure as the evaluation network, and establish the update frequency for the target network.
\subsubsection{Establish the Experience Replay Mechanism}
Initialize a replay memory to store experience samples $(S_t, A_t, R_t, S_{t+1})$ generated during the agent's interaction with the environment. These samples consist of the current state $S_t$, the action $A_t$ taken, the reward $R_t$ received, and the next state $S_{t+1}$ reached after performing the action.
\subsubsection{Dynamic Oracle Node Selection}
\begin{itemize}
    \item Execute Action and Update Strategy: At each time step $t$, an action $A_t$ is chosen based on the current state $S_t$ using the $\epsilon-greedy$ strategy. Specifically, with a probability of $1- \epsilon$, the action that maximizes Q-value is selected, which corresponds to choosing the oracle with the highest expected reward. With a probability of $\epsilon$, an action is randomly selected to explore new potential optimal strategies. The chosen action $A_t$ (selecting an oracle) is then executed, an immediate reward $R_t$ is received, and the next state $S_{t+1}$ is observed.
    \item Experience Replay and Learning: The current experience is stored in the replay memory. Then, updates are performed using experience replay: a small batch of samples $(S_k, A_k, R_k, S_{k+1})$ is randomly drawn from the replay memory for training. The target Q-value ${target}_k$ is calculated as follows:
    \begin{equation}
    {target}_k = r_k + {\gamma}\operatorname{max}_{A^{\prime}} {\hat{Q}}\left(S_{k + 1}; A^{\prime}; {\theta}^{\prime}\right)
    \end{equation}
    Where $\hat{Q}$, ${\theta}^{\prime}$ represent the output and parameter of the target network, respectively.  
    Next update the parameter $\theta$ of the evaluation network by minimizing the loss function:
    \begin{equation}
    \mathcal{L}(\theta) = \mathbb{E} \left[ \left( {target}_k - Q(S_k, A_k; \theta) \right)^2 \right]
    \end{equation}
    \item Update Target Network: Every fixed number of steps, copy the parameter $\theta$ of the evaluation network to the parameter ${\theta}^{\prime}$ of the target network:
    \begin{equation}
        \theta^{\prime} \leftarrow \theta
    \end{equation}
\end{itemize}
\subsubsection{Iterative Training Process}
The algorithm repeatedly executes step 3) until the defined stopping condition is reached, such as achieving a certain level of convergence or reaching a specified number of training iterations.

\begin{algorithm}[!t]   
  \caption{Dynamic Oracle Node Selection Algorithm in TCO-DRL Based on DQN}  
  \label{alg:1}  
  \begin{algorithmic}[1]  
    \State \textbf{Input:} initial $ \epsilon $, $ \alpha $, $ \gamma $, learning frequency $ f $, minibatch $ S_{\Delta} $, replay period $ \eta $
    \State Initialize replay memory $ \Delta $ with capacity $ N_{\Delta} $
    \State Initialize evaluation value function $ Q $ with random parameters $ \theta $
    \State Initialize target value function $ \hat{Q} $ with random parameters $ {\theta}^{\prime} $
     \For{each new request $ i $ arrives at $ t $}
     \State with probability $ \epsilon $, randomly choose an action; otherwise $ A_t=\operatorname{argmax}_A Q\left(S_t ; A ; \theta\right) $
     \State Dispatch request $ i $ according to action $ A_t $, receive reward $ R_t $, and observe state transition at next decision time $ t+1 $ with a new state $ S_{t+1} $
     \State Store transition $\left(S_t, A_{t}, R_{t}, S_{t+1}\right) $ in $ \Delta $
      \If {$ j $ $\geq$ $ t $ {and} $ t $ $ \equiv $ 0 mod $ f $}  
		\If {$ j $ $ \equiv $ 0 mod $ \eta $}
     	   \State Reset $ {\hat{Q}} = {Q} $
             \EndIf
             \State randomly select samples $ S_{\Delta} $ from $ N_{\Delta} $
             \For{each transition $\left(S_k, A_{k}, R_{k}, S_{k+1}\right) $ in $ S_{\Delta} $ }
                \State $ {target}_k = r_k + {\gamma}\operatorname{max}_{A^{\prime}} {\hat{Q}}\left(S_{k + 1}; A^{\prime}; {\theta}^{\prime}\right)$
                \State update DNN parameters $\theta$ with loss function
  			 \EndFor
            \State Gradually decrease $\epsilon$ until to the lower bound
		\EndIf   
  	\EndFor \\
        \Return action $ A_i $
  \end{algorithmic}  
\end{algorithm}

\subsubsection{Optimal Oracle Selection Strategy}
\begin{figure*}[!t]
\centering
\includegraphics[width=0.9\textwidth]{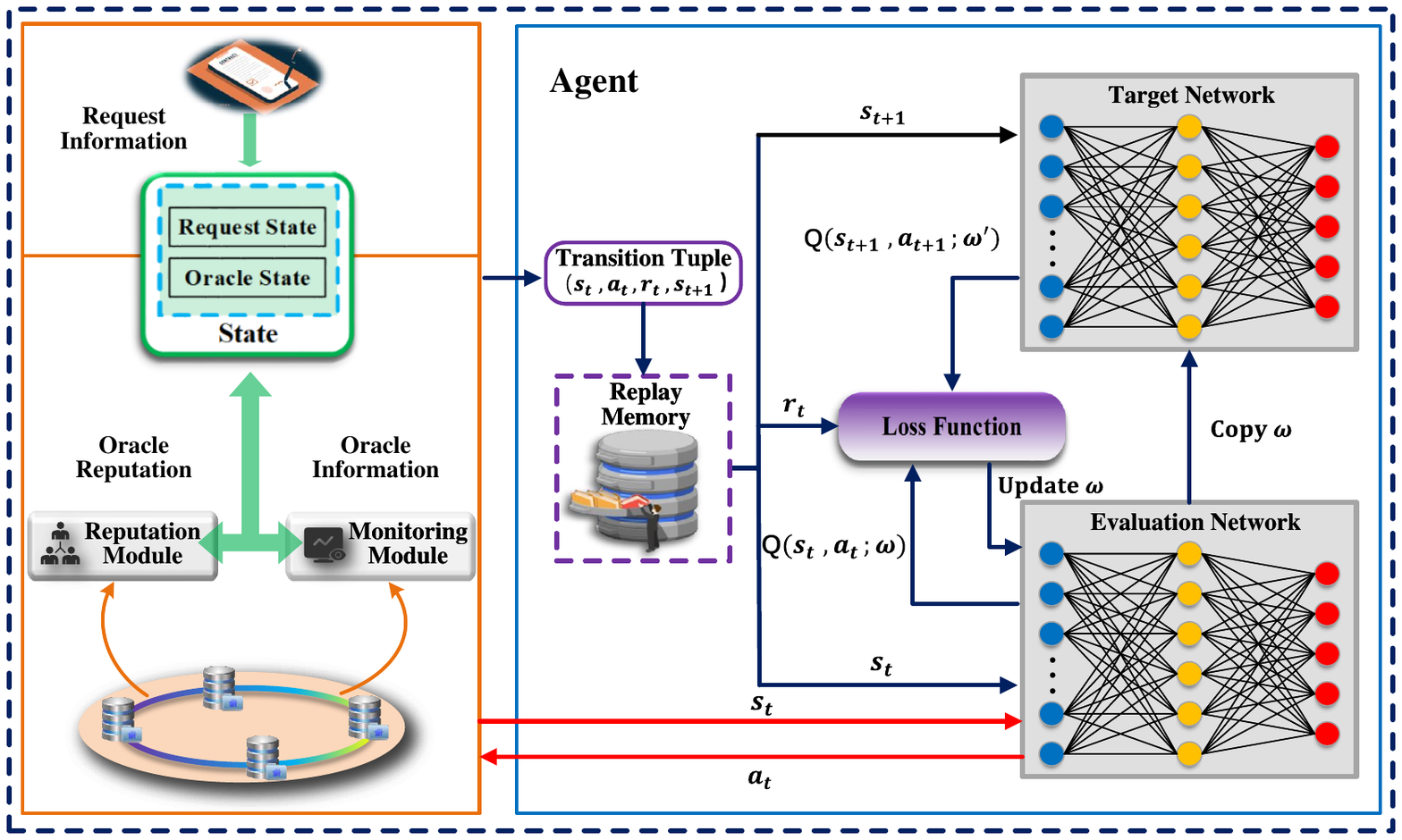}
\caption{Oracle node selection process based on DQN.}
\label{fig_4}
\end{figure*}

\begin{itemize}
    \item Policy Evaluation and Improvement: The training process of the DQN is represented by the dark blue lines in the Fig. \ref{fig_4}. After training is completed, the agent, using the learned Q-network, can select the optimal oracle in a given state to maximize cumulative rewards.
    \item Policy Implementation: The actual execution process of the DQN is represented by the red lines in the Fig. \ref{fig_4}. By applying the learned policy to the real-world oracle selection task, it is possible to dynamically choose the most suitable oracle based on real-time state information, such as the current reputation and cost of the oracles.
\end{itemize}
The overall training and execution process of the algorithm is summarized in Algorithm \ref{alg:1}.

\section{Simulation Results and Evaluations}
In this section, we provide a detailed analysis of the experimental design and evaluation results. We begin by describing the configuration of the experimental platform, the baseline methods for comparison and the simulated attacks. Next, we compare the robustness and performance of TCO-DRL with other baseline methods in environments with noise and progressively worsening conditions. Finally, we deploy TCO-DRL on blockchain and simulate three classic attacks to test its effectiveness.
\subsection{Simulation Settings}
\subsubsection{Platform Settings}
As described in the work \cite{taghavi2023reinforcement}, there is currently no available blockchain oracle dataset. The study work \cite{taghavi2023reinforcement} used synthetic data for experiments, and we have referenced its data on cost distribution. Additionally, based on the work \cite{taghavi2023reinforcement}, we have adjusted the parameters of the experimental data to closely match the real-world operation of oracle communities. Specifically, we designed the complexity of data requests and the performance metrics of oracles based on the average response time of data requests in BandChain \cite{bandprotocol}, making it to 6 seconds to align with BandChain's average. The arrival time of data requests follows a Poisson distribution. Regarding the pricing model, we adopted a fixed-rate pricing strategy, similar to most oracle communities, where a fixed fee is charged for each transaction or request.

The experiment simulated an oracle community consisting of 15 oracles, categorized into three types based on the service requests they can handle (e.g., data update frequency), with five oracles of each type. A total of 6,000 data requests were generated. The oracle community was simulated using Python, with configurations of Python 3.6 and TensorFlow 2.0.0. We set up a blockchain network using Ethereum on Ubuntu 18.04 and deployed and tested the smart contracts using the Truffle framework. All experiments were conducted on a host equipped with a 2.5 GHz Intel Core i5-7300HQ CPU and 8 GB RAM. The specific experimental parameters are shown in the Table \ref{tab:4}.

\begin{table}[!t]
\centering
\caption{SYSTEM SIMULATION PARAMETERS}
\label{tab:4}
\begin{tabular}{ll}
\toprule
\textbf{Parameter} & \textbf{Value} \\
\midrule
Capacity of reply memory $ N_{\Delta} $ & 800 \\
Capacity of minibatch $ S_{\Delta} $ & 30\\
Learning rate $ \alpha $ & 0.01 \\
Discount factor $ \gamma $ & 0.9 \\
Number of oracles & 15 \\
Number of data requests & 6000 \\
Data requests complexity & ${\mathcal N} \left(6000, 500\right) $  \\
Length of the sliding time window & 5\\
Initial reputation value & 0.5\\
Trust threshold & -1.5\\
Weights of calculating reliability score &  \makecell[tl]{$\omega$ = 0.2 $\varphi$ = 0.4 \\$\psi$ = 0.4} \\
The harm scores for the four behavior categories & \{0,1,5,100\} \\
Weights of calculating base reputation score & \makecell[tl]{$\xi$ = 0.4 $\zeta$ = 0.4 \\$\delta$ = 0.2}\\
Weight of time factor $\chi$ & 0.6 \\
Weights of calculating reward & \makecell[tl]{$\vartheta$ = 2.5 $\lambda$ = 1.5 \\ $\mu$ = 4}\\
\bottomrule
\end{tabular}
\end{table}

\subsubsection{Baseline Methods}
TCO-DRL was experimentally compared with a classic baseline method, Round-Robin, as well as two recent baseline methods: BLOR and PSG.
\paragraph{Round-Robin \cite{rasmussen2008round}}
Round-Robin is a commonly used allocation algorithm that is simple and easy to implement, primarily designed for the fair distribution of resources. This method assigns requests in a fixed order, ensuring that each oracle node has an equal opportunity to provide services.
\paragraph{BLOR \cite{taghavi2023reinforcement}}
BLOR is a method that uses a reinforcement learning approach called Bayesian Multi-Armed Bandit algorithm to intelligently select oracle nodes. It is the first smart mechanism specifically designed for oracle node selection and serves as one of the main baseline methods for our comparison. BLOR aims to optimize both the cost and reputation of the oracles. It is important to note that BLOR evaluates reputation based on the number of successful and failed requests handled by the oracles. The original study generated the corresponding historical records through simulation, but we did not have access to such data. To address this initialization issue for BLOR, we employed Round-Robin to fairly assign data requests during the initial training phase (for a short period) to generate historical records for the oracles.
\paragraph{PSG \cite{azizi2022deadline}}
PSG is a semi-greedy method originally used for task scheduling in the previous work. Essentially, it is a decision optimization method, so we have applied and implemented its concept in the selection of oracle nodes. First, we filter out all oracles that would receive a reward greater than zero after executing a specific data request. Then, we sort these oracles in ascending order based on cost, and finally, a random oracle is selected from the top $q$ oracles to handle the data request.

\subsubsection{Simulated Attacks}
We referred to the work \cite{marche2020trust} to simulate three common trust-related attacks for IoT.
\paragraph{Malicious With Everyone (ME)}
Malicious nodes exhibit harmful behavior towards all requesters. This is a basic type of attack where the node always provides low-quality service and unreliable recommendations, regardless of who the requester is.
\paragraph{On-Off Attack (OOA)}
Nodes periodically switch their behavior between honest (on) and malicious (off) states. In the honest phase, the node establishes trust, while in the malicious phase, it exploits this trust to launch attacks.
\paragraph{Opportunistic Service Attack (OSA)}
Malicious nodes provide high-quality service only when they notice their trust reputation declining. This strategy aims to maintain a satisfactory trust level, ensuring the node still has opportunities to provide services.

\subsection{Performance Comparison}
\begin{figure}[!t]
\centering
\includegraphics[width=3in]{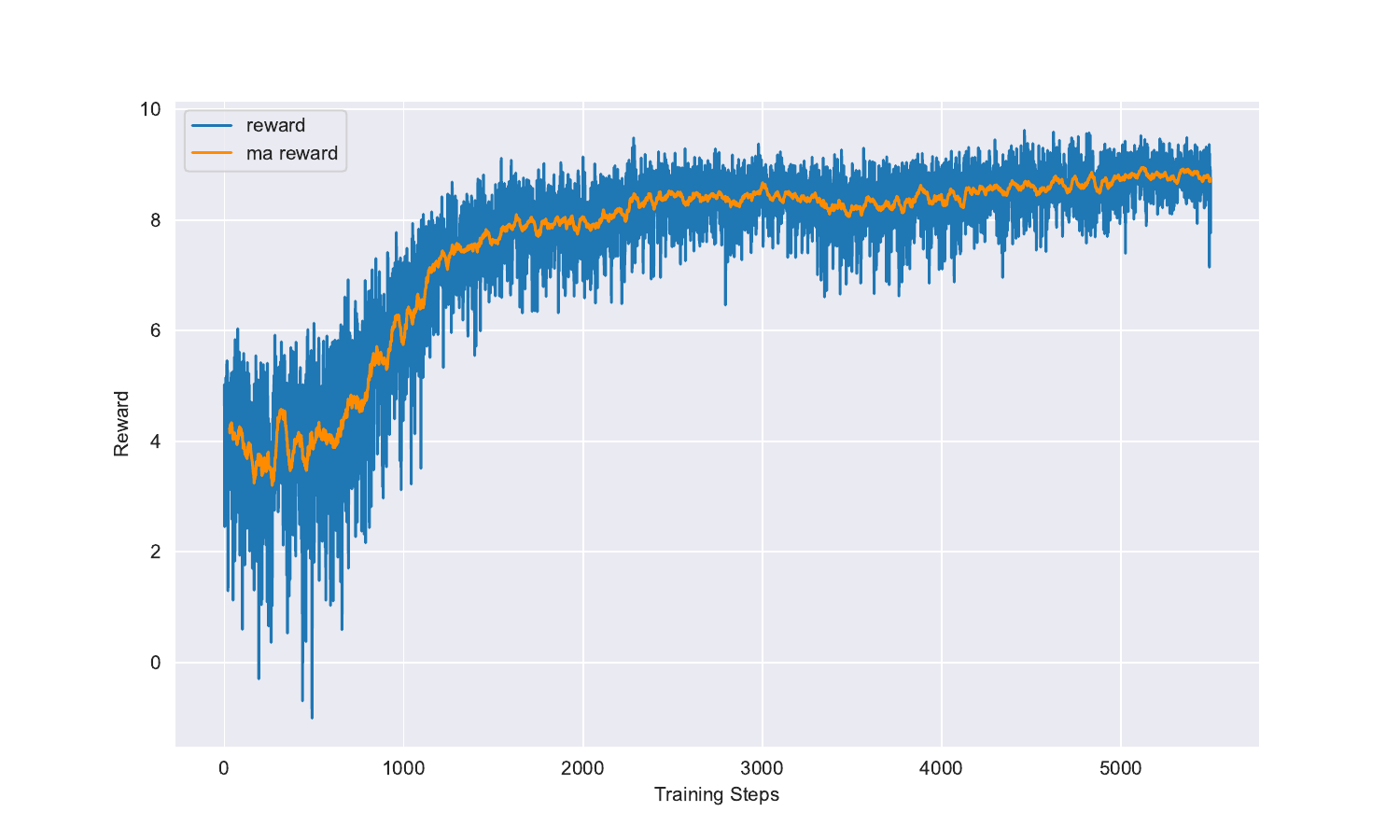}
\caption{Convergence performance of TCO-DRL.}
\label{fig_5}
\end{figure}

We classify oracles based on their behavior into three categories: trusted oracles, benign oracles, and malicious oracles. Trusted oracles are capable of providing high-quality and reliable services over the long term. Benign oracles are generally honest and reliable but may occasionally make mistakes, with performance and data quality that are not as high as those of trusted oracles. Malicious oracles intentionally provide false or misleading data to manipulate the system or harm other participants. In our community setup, we included three malicious oracles and three benign oracles, with the remaining oracles being trusted. It should be noted that, to better test the robustness of methods, we did not set a trust threshold for oracles in the comparative experiments. This means that even if an oracle engages in malicious behavior, causing its reputation value to fall below the trust threshold, it can still be selected to provide services. This design enables us to evaluate the performance of each method under the influence of persistent malicious behavior by oracles. Fig. \ref{fig_5} shows the convergence of TCO-DRL, with the model reaching convergence after approximately 2,000 training steps.

\begin{figure}[!t]
\centering
\includegraphics[width=2.5in]{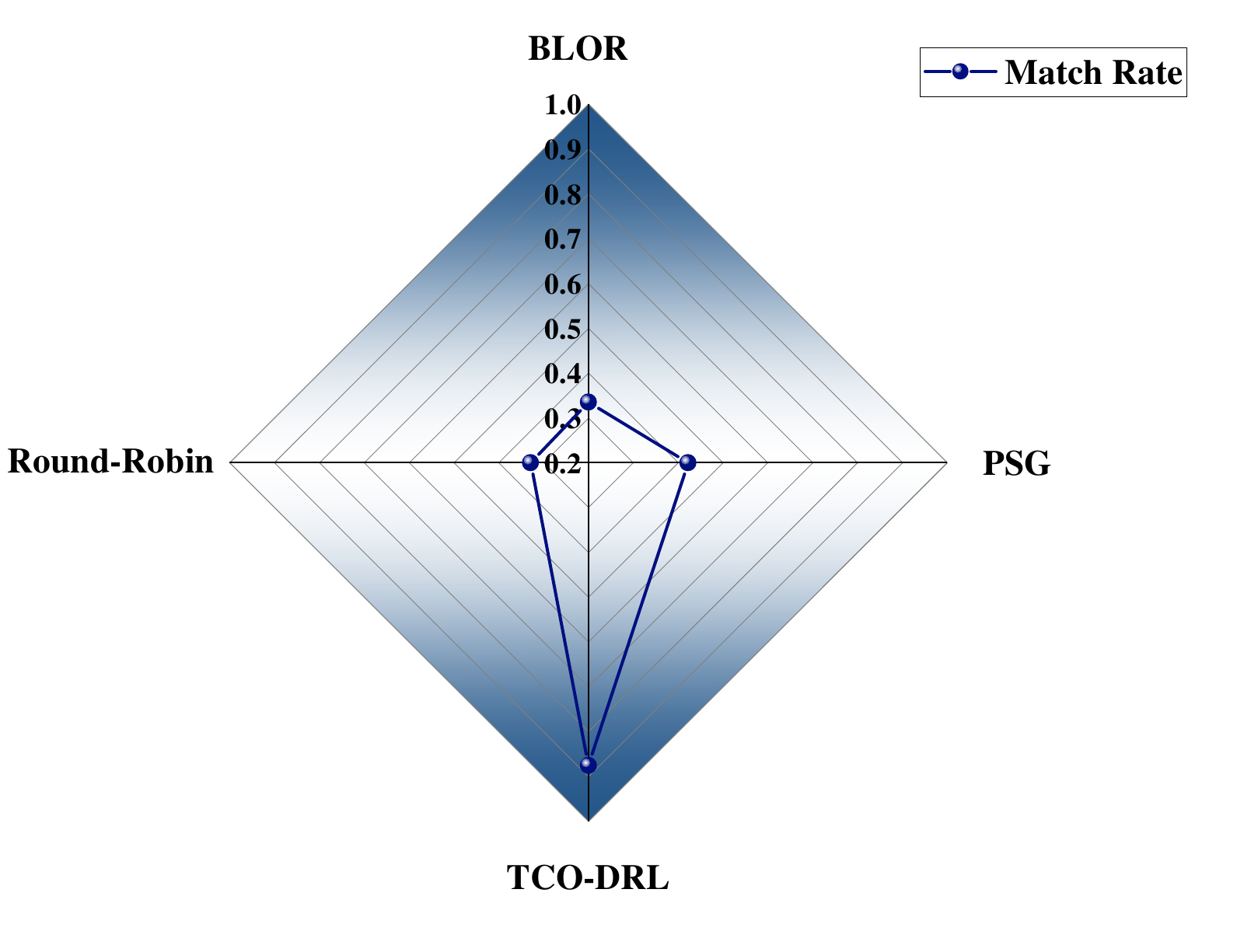}
\caption{Match rate comparison of different methods.}
\label{fig_6}
\end{figure}

Fig. \ref{fig_6} shows the matching results of different methods for assigning requests to oracles. TCO-DRL has a significant advantage, successfully matching the services required by most requests (87.5\%) with those provided by the oracles. In contrast, Round-Robin assigns requests in a fixed order, resulting in completely random matches. BLOR shows poorer matching results because its source code does not provide corresponding optimization objectives. PSG performs better than Round-Robin and BLOR since it filters oracles with a reward value greater than zero. However, because the reward function emphasizes reputation and cost, PSG is limited when faced with more than two optimization objectives, resulting in a significantly lower matching rate compared to TCO-DRL.

\begin{figure}[!t]
\centering
\includegraphics[width=2.5in]{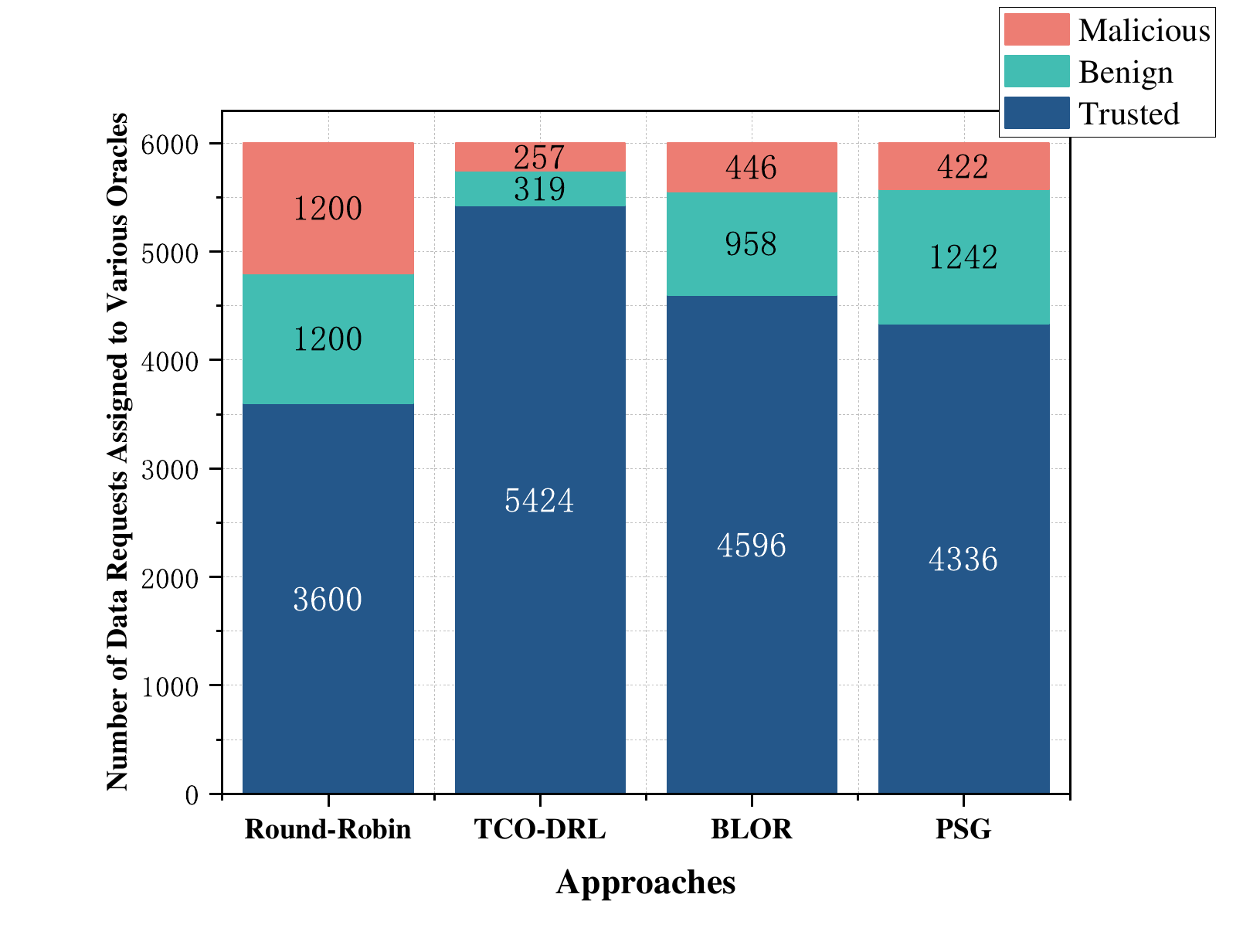}
\caption{Distribution of data requests assigned to various oracles using different methods.}
\label{fig_7}
\end{figure}

 Fig. \ref{fig_7} shows the distribution of data requests assigned to oracles with different behaviors using various methods. It is evident that TCO-DRL significantly reduces the number of requests assigned to malicious oracles: only 4.28\% of requests are assigned to malicious oracles with TCO-DRL, which is a reduction of 39.10\% to 78.58\% compared to other methods. This indicates that TCO-DRL excels at identifying and avoiding malicious oracles, effectively reducing the interference of malicious nodes in the system. Additionally, the proportion of requests allocated to trusted oracles by TCO-DRL is as high as 90.40\%, which is an increase of 15.27\% to 33.63\% over other methods. Moreover, the number of requests assigned to benign oracles by TCO-DRL is noticeably lower than that by other methods. These results indicate TCO-DRL's sensitivity to oracle reputation values and its tendency to prioritize selecting oracles with good behavior—even though benign oracles have a higher reputation than malicious ones and are a reasonable choice, TCO-DRL always strives to allocate requests to oracles with a highest reputation.

\begin{figure}[!t]
\centering
\includegraphics[width=2.5in]{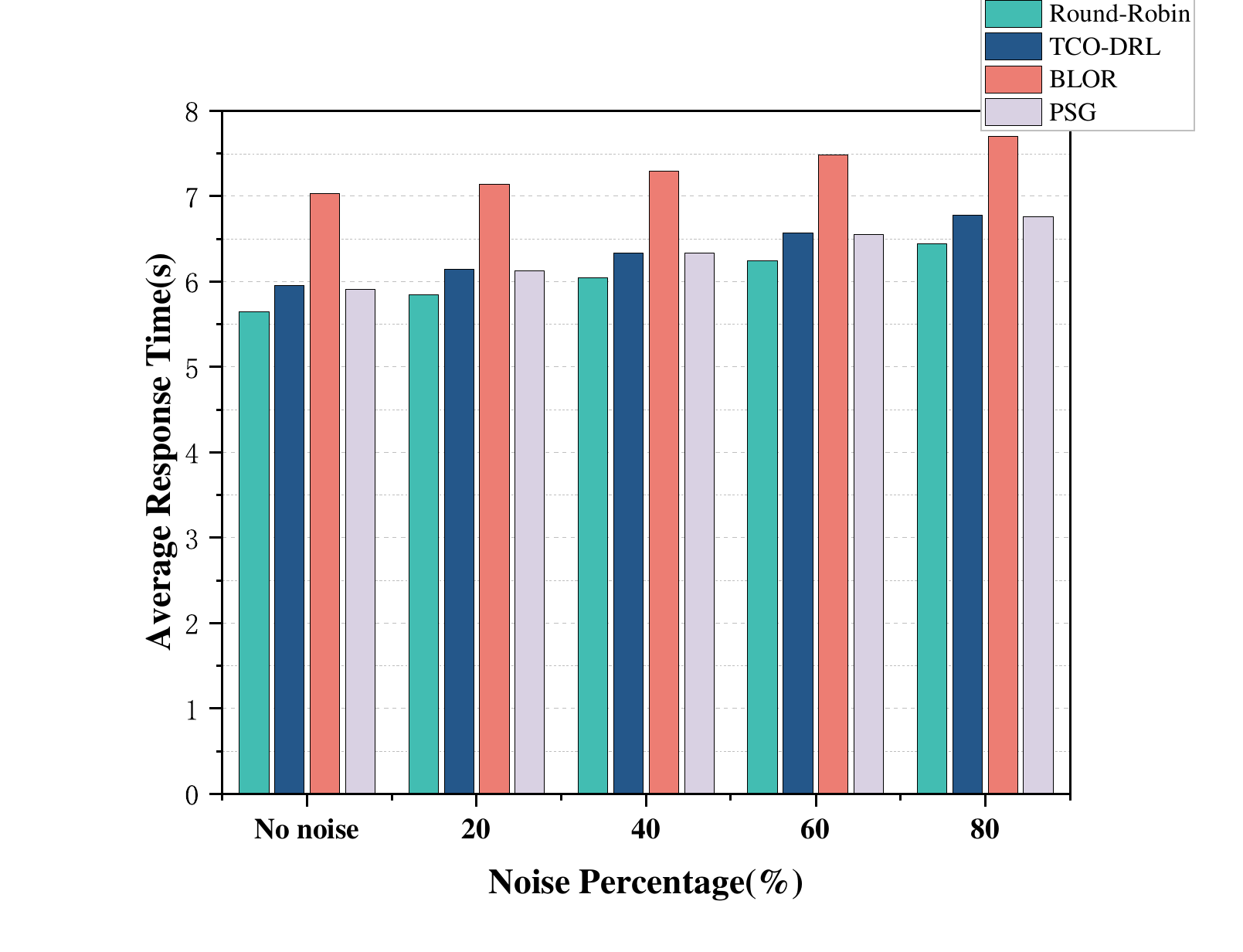}
\caption{Average response time comparison of different methods against noise observations.}
\label{fig_8}
\end{figure}

We then tested the performance of the four methods in noisy environments. Noise refers to situations where the behavior of oracles does not occur as expected. Fig. \ref{fig_8} shows the average response time of different methods as noise increases. It can be observed that the average response time for all methods increases with the increase in noise. Round-Robin has a relatively lower average response time because its algorithm is simple and straightforward, allowing for quicker decision-making. TCO-DRL and PSG show relatively good average response times across all noise percentages, with more stable variations, indicating better robustness. BLOR has the longest average response time, possibly because it always selects oracles with high reputation scores and low costs, leading to increased waiting times for data requests and thus raising the average response time.

\begin{figure}[!t]
\centering
\includegraphics[width=2.5in]{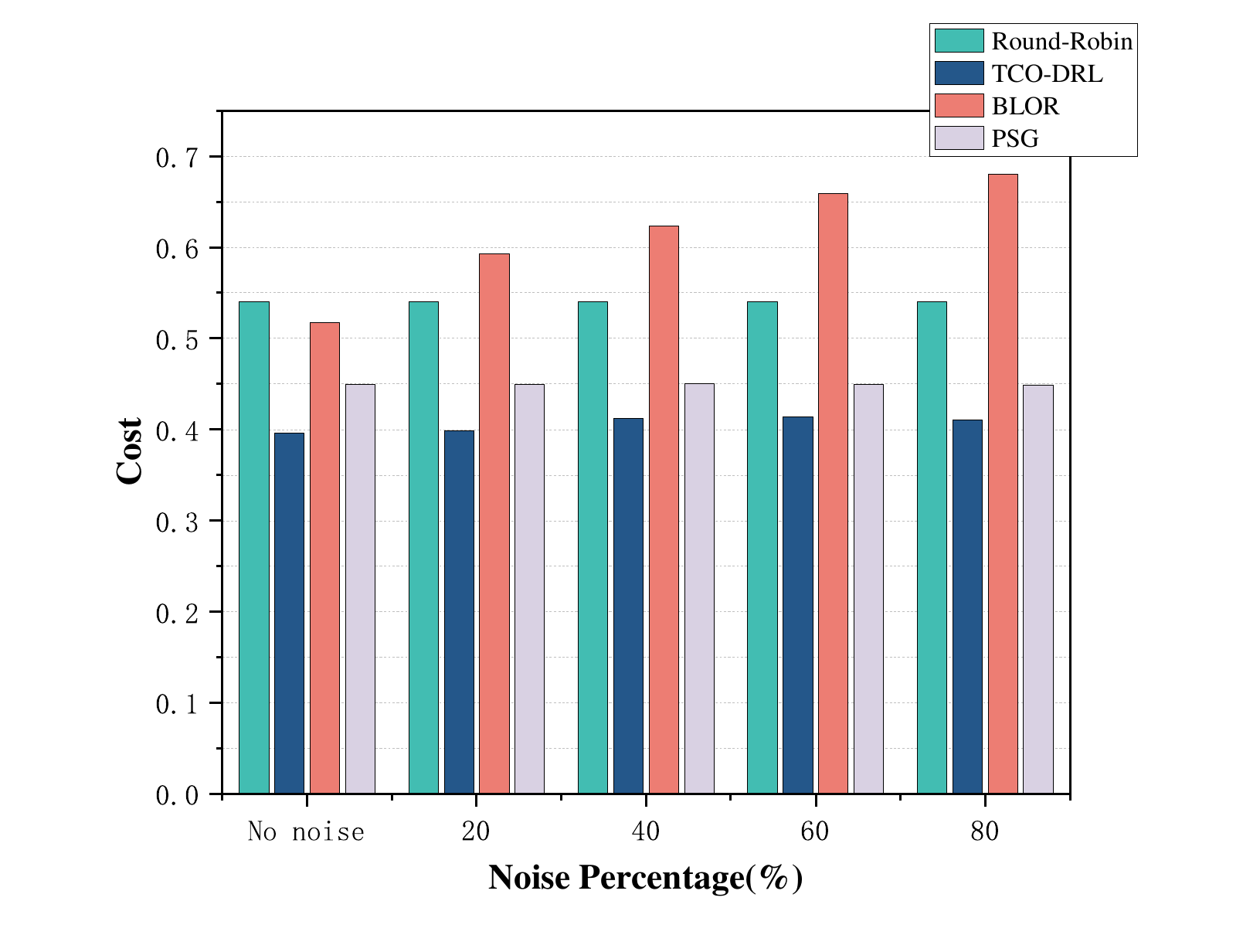}
\caption{Cost comparison of different methods against noise observations.}
\label{fig_9}
\end{figure}

Next, we examined how the cost changes with noise for different methods, as shown in Fig. \ref{fig_9}. It can be seen that the costs for Round-Robin, TCO-DRL, and PSG remain stable, suggesting strong resistance to noise interference. In contrast, the cost for the BLOR method varies significantly with noise, indicating a higher sensitivity to noise. The reason for the less noticeable cost variation curve in Round-Robin is the fixed-rate pricing model and its fixed order for request allocation. TCO-DRL and PSG exhibit more stable cost changes because their algorithms are better adapted to environmental changes, providing greater robustness. BLOR's poorer performance might be due to greater interference from noise in its decision-making process, a result consistent with the findings in the work \cite{taghavi2023reinforcement}.

\begin{figure}[!t]
\centering
\includegraphics[width=2.5in]{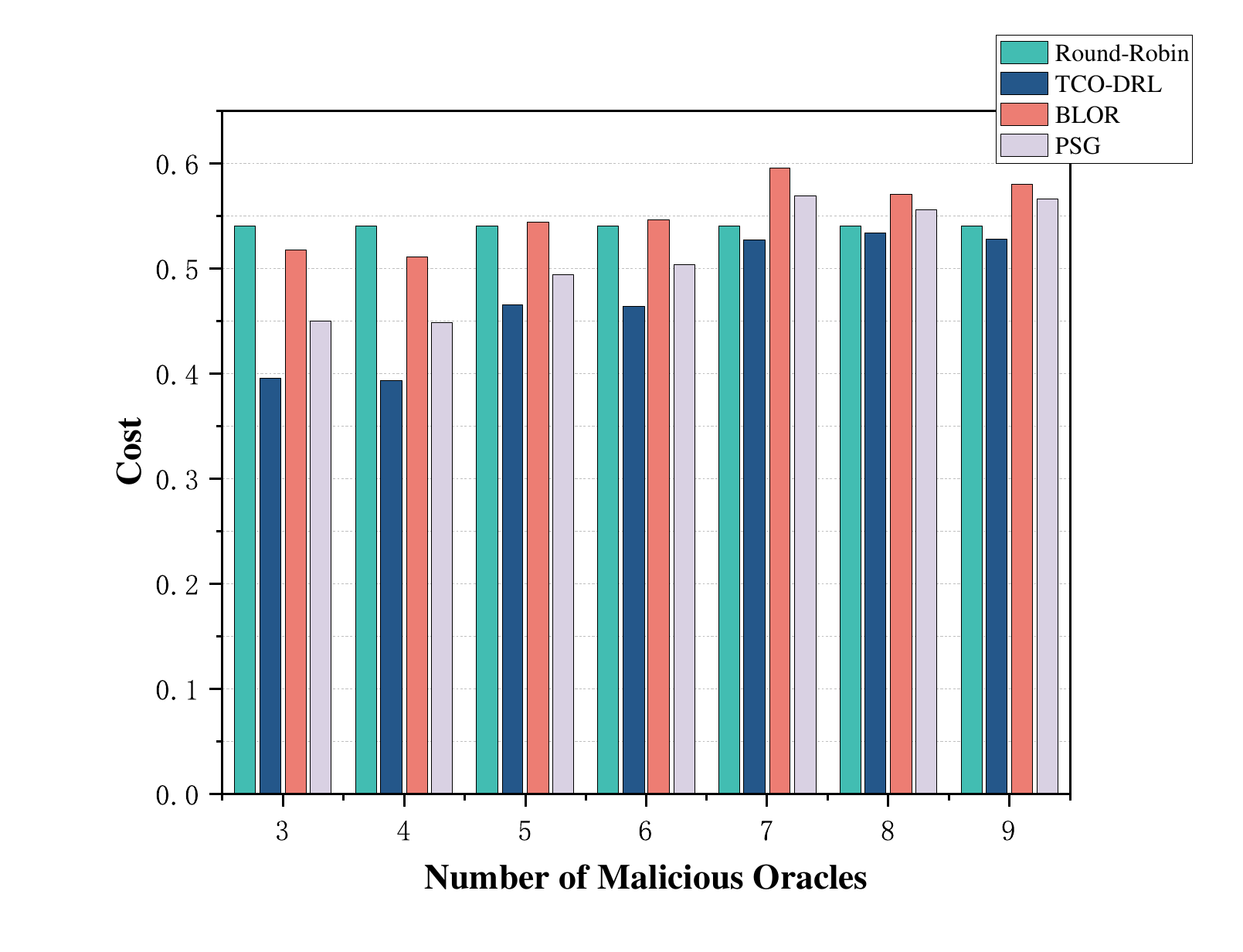}
\caption{Cost comparison of different methods with increasing malicious oracles.}
\label{fig_10}
\end{figure}

Finally, we tested the robustness of the four methods in increasingly adverse environments. Each time, we randomly selected one oracle from the remaining benign and trusted oracles to turn it into a malicious oracle, and we analyzed the changes in cost and the distribution of data requests for each method. Fig. \ref{fig_10} illustrates the cost variations across different methods. When there are three malicious oracles in the community, the average costs for Round-Robin, BLOR, and PSG are 0.540, 0.518, and 0.450, respectively, while TCO-DRL achieves a significantly lower average cost of 0.396, reducing costs by 12.00\% to 26.67\% compared to the other methods. This demonstrates that TCO-DRL is more effective at selecting low-cost oracles. As the number of malicious oracles in the community increases, the costs for TCO-DRL, BLOR, and PSG also rise. This is because all three algorithms strive to select oracles with high reputation values to handle data requests, but oracles with higher reputation values typically incur higher costs, thereby increasing the average cost. Notably, even when the number of malicious oracles in the community reaches nine (more than half of the total), TCO-DRL maintains a cost of 0.528, still below the overall average cost of 0.54 for all oracles. This indicates that TCO-DRL has a significant advantage in optimizing average costs.

\begin{figure}[!t]
\centering
\includegraphics[width=2.5in]{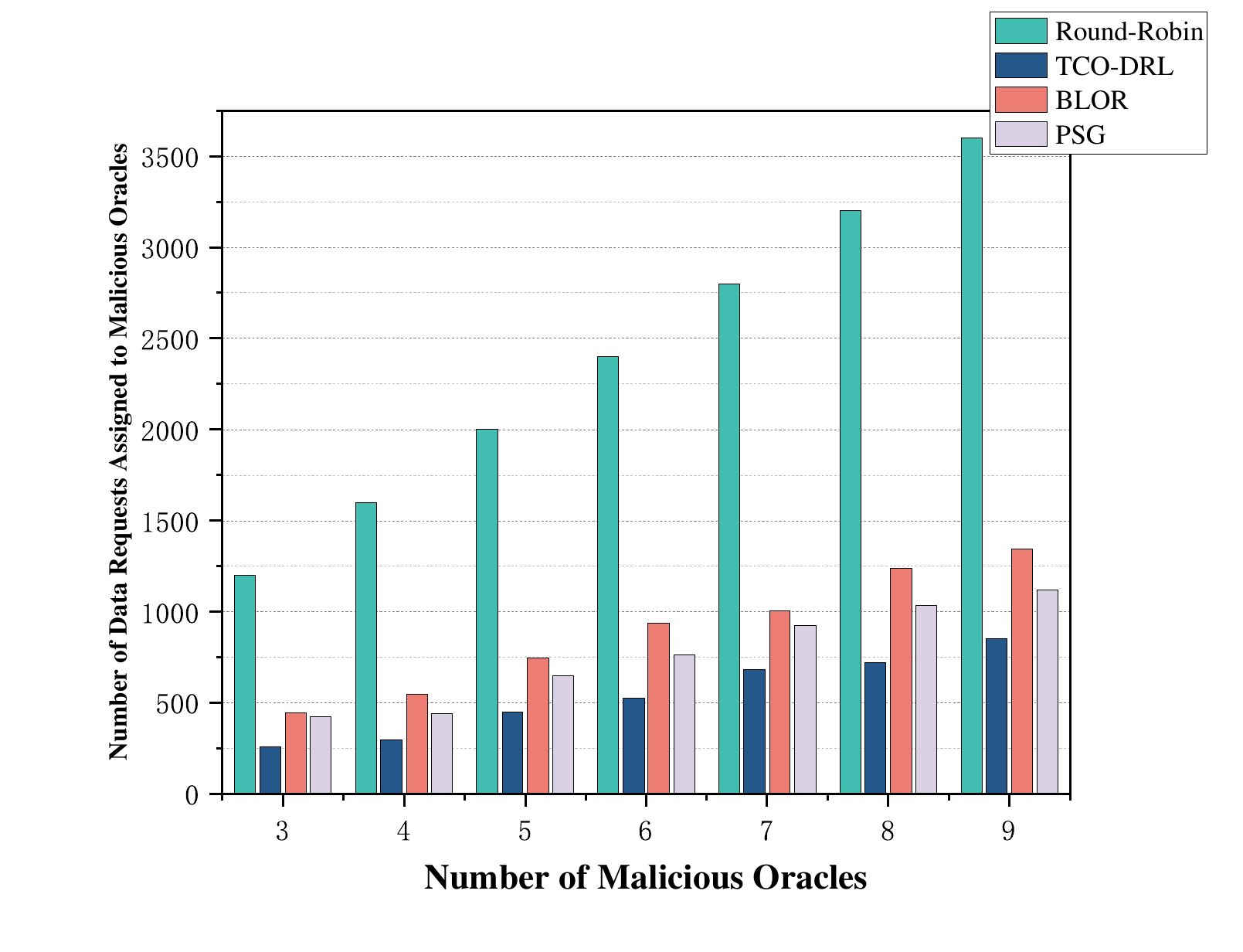}
\caption{Comparison of data requests assigned to malicious oracles with increasing malicious oracles.}
\label{fig_11}
\end{figure}

\begin{figure}[!t]
\centering
\includegraphics[width=2.5in]{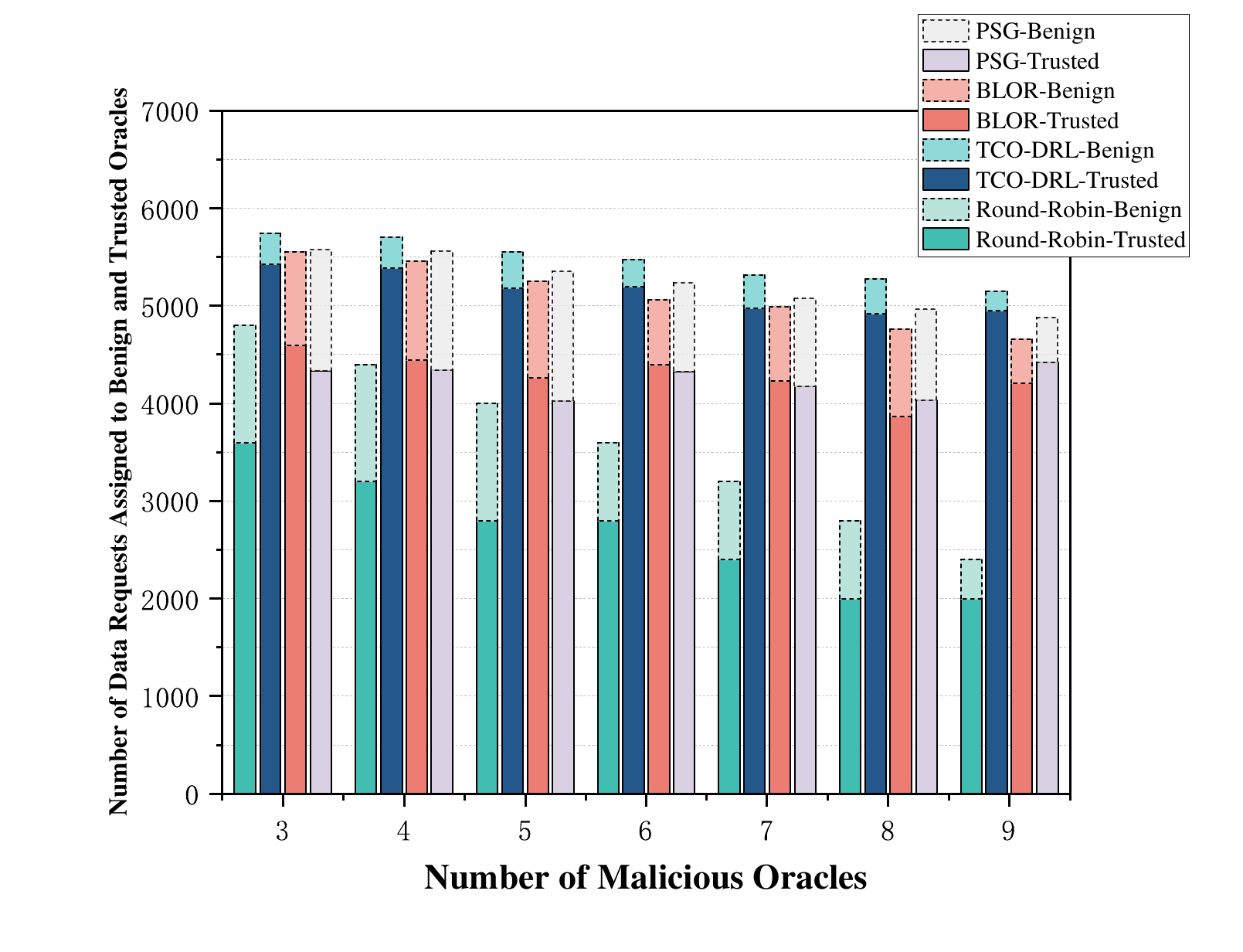}
\caption{Comparison of data requests assigned to benign and trusted oracles with increasing malicious oracles.}
\label{fig_12}
\end{figure}

Fig. \ref{fig_11} depicts the changes in the number of data requests assigned to malicious oracles by each method. As the number of malicious oracles increases, the number of data requests assigned to them by all methods also rises. When the number of malicious oracles reaches nine, BLOR and PSG assign 1,342 and 1,118 requests, respectively, far below the theoretical average of 3,600, indicating their effectiveness in selecting trusted oracles. TCO-DRL assigns only 852 requests to malicious oracles, reducing the number by 23.79\% to 76.33\% compared to the other three methods, demonstrating superior capability in identifying malicious oracles and adapting to different environments. Additionally, Fig. \ref{fig_12} shows the number of requests assigned to benign and trusted oracles by each method. Compared to the other three methods, TCO-DRL not only assigns the highest total number of requests to benign and trusted oracles but also has the smallest proportion of requests assigned to benign oracles, ranging from 3.77\% to 6.74\%. In contrast, the proportions for Round-Robin, BLOR, and PSG are 16.66\% to 30.00\%, 9.62\% to 18.84\%, and 9.46\% to 24.86\%, respectively. This indicates that TCO-DRL not only has a finer ability to differentiate between nodes but also shows more stable variations than the other three methods, demonstrating stronger robustness.

In summary, the experiments demonstrate TCO-DRL's clear advantages in selecting trusted and cost-effective oracles, while also showing strong resistance to interference and robustness.

\subsection{Effectiveness Analysis}
\begin{figure}[!t]
\centering
\includegraphics[width=2.5in]{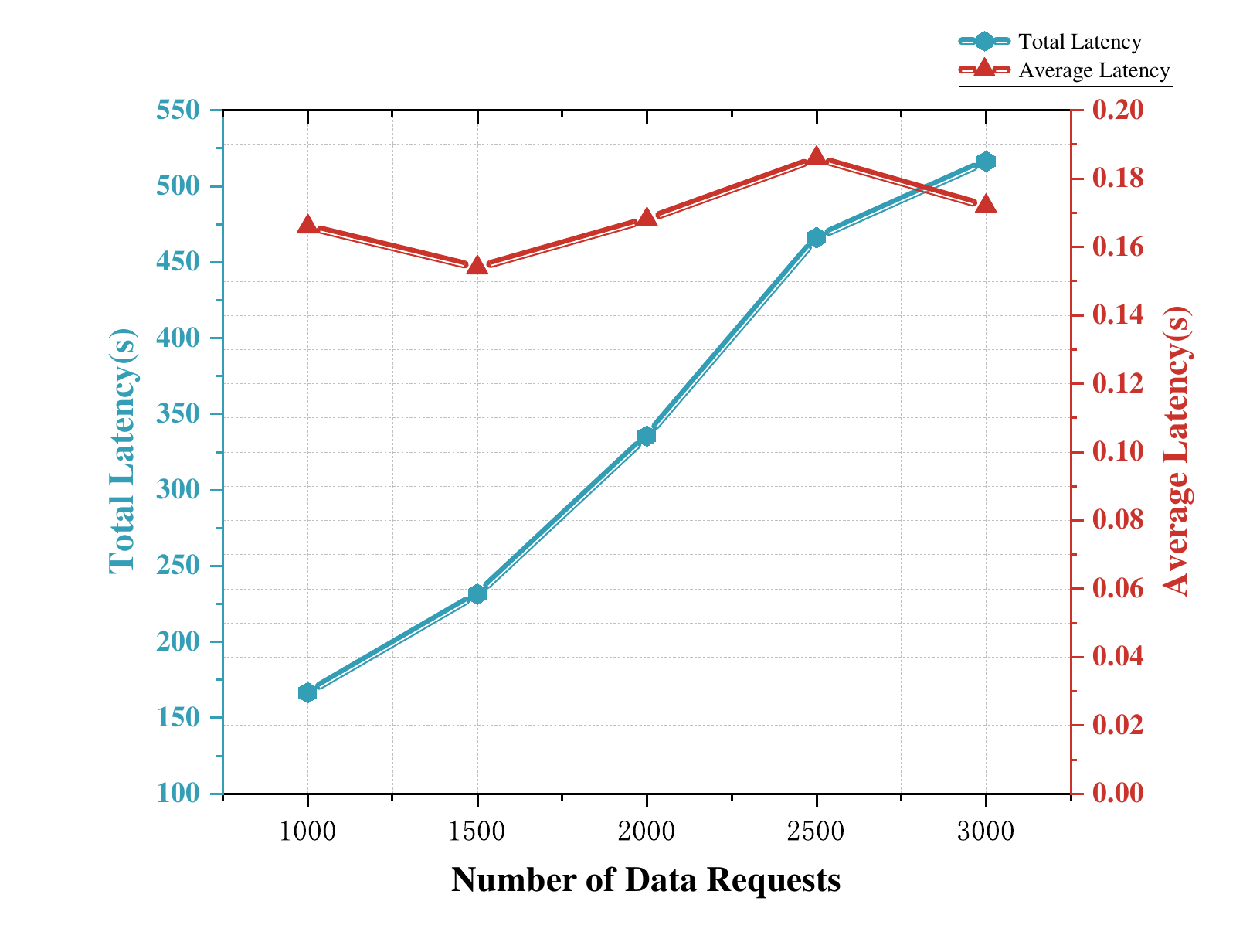}
\caption{Total and average latency with increasing data requests.}
\label{fig_13}
\end{figure}

\begin{figure}[!t]
\centering
\includegraphics[width=2.5in]{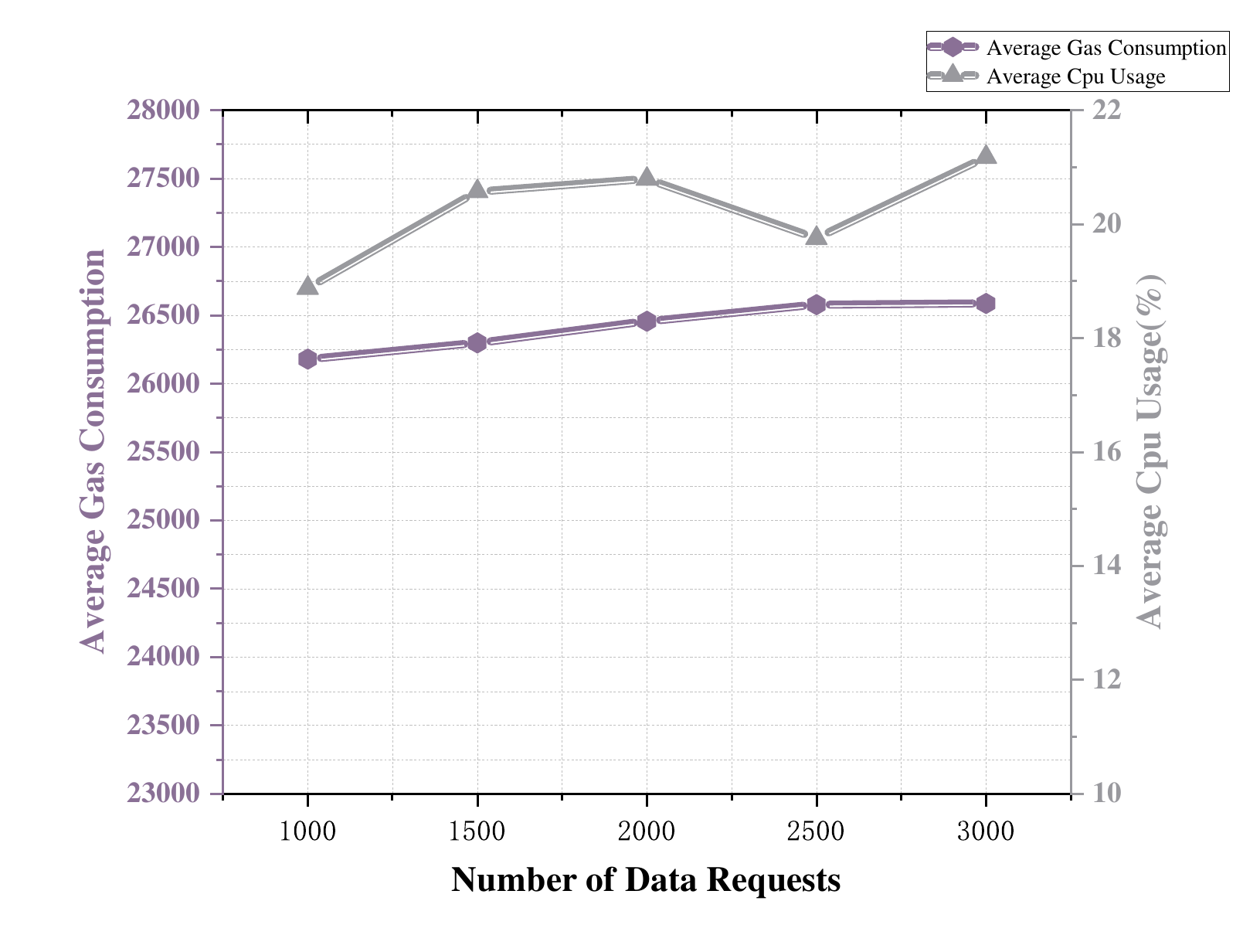}
\caption{Gas consumption and CPU usage with increasing data requests.}
\label{fig_14}
\end{figure}

In this part of the experiment, we deployed TCO-DRL on Ethereum to test its performance and resistance to attacks. 
Fig. \ref{fig_13} shows the changes in total latency and average latency of TCO-DRL. Here, latency refers to the time required from initiating a transaction to receiving a transaction receipt, i.e., the time needed for transaction confirmation. As the number of requests increases, the system latency fluctuates but generally remains stable, with the average latency consistently below 0.2 seconds. We then tested the average gas consumption for contract invocations and the average CPU usage of TCO-DRL, as shown in Fig. \ref{fig_14}. As the number of data requests increases, the average gas consumption remains stable and consistently low. This is due to the hybrid on-chain and off-chain execution architecture, which moves complex computation off-chain, thereby avoiding additional gas costs associated with code complexity. Additionally, the CPU usage, which stays around 20\%, demonstrates that the algorithm can effectively handle more data requests without significantly increasing resource consumption.

We then evaluated the performance of TCO-DRL when facing three simulated attacks. It is important to note that in this part of the experiment, we set a trust threshold: oracles with a reputation value below this threshold will not be allowed to provide services until their reputation recovers above the threshold. Our trust management scheme permits an oracle to commit up to one instance of moderate harm, so we set the trust threshold at -1.5.

\begin{figure}[!t]
\centering
\includegraphics[width=2.5in]{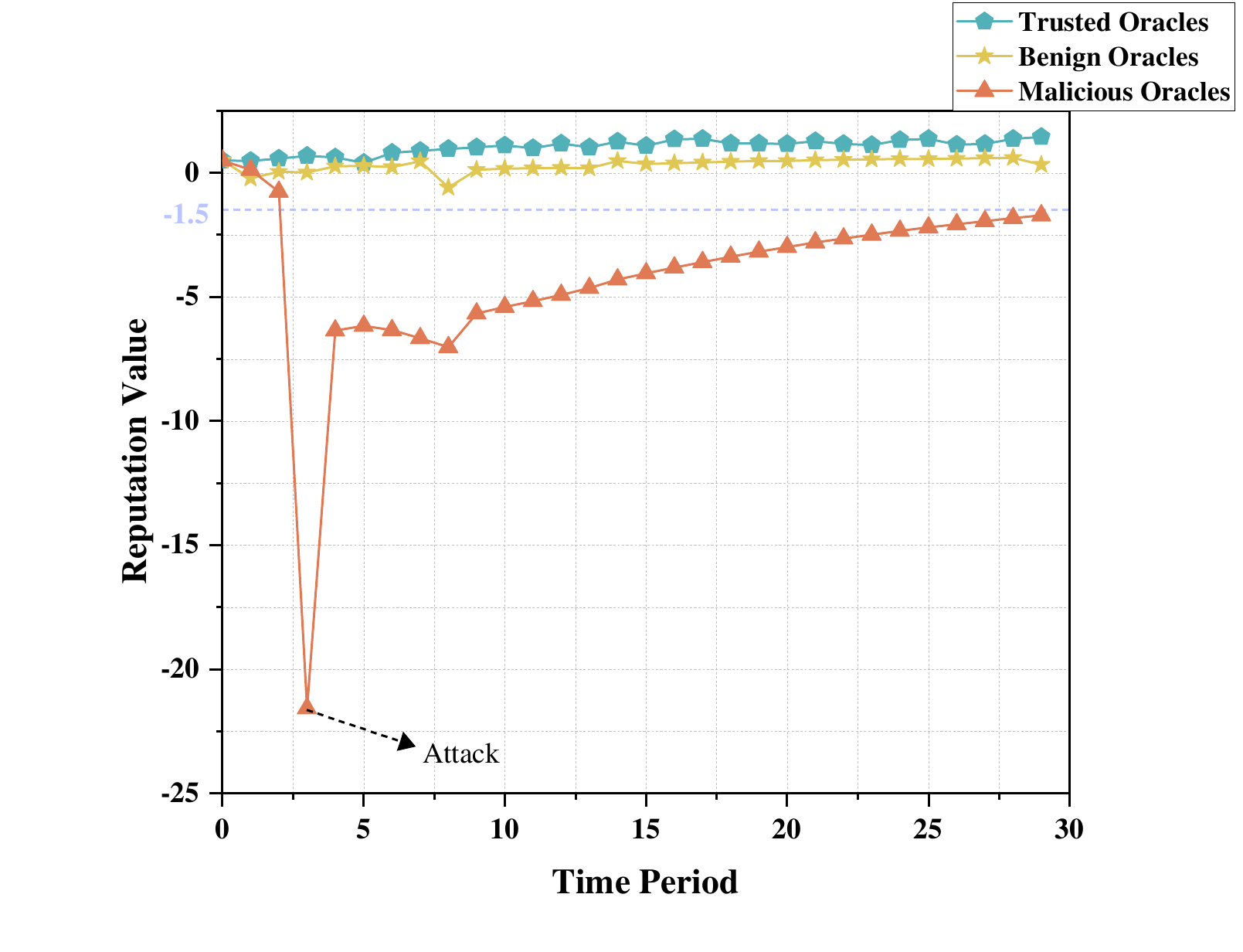}
\caption{Change in reputation value under malicious with everyone attack.}
\label{fig_15}
\end{figure}

We first tested TCO-DRL's resistance to the malicious with everyone (ME) attack. As shown in Fig. \ref{fig_15}, our trust management scheme can clearly differentiate between the types of oracles based on their reputation values, indicating its effectiveness in accurately capturing each oracle's behavior and mapping it to a reputation score. Once a malicious node initiates an ME attack, its reputation value rapidly declines below the trust threshold, resulting in it being prohibited from providing services. The recovery of its reputation is a slow process, effectively preventing the malicious node from launching further ME attacks.

\begin{figure}[!t]
\centering
\includegraphics[width=2.5in]{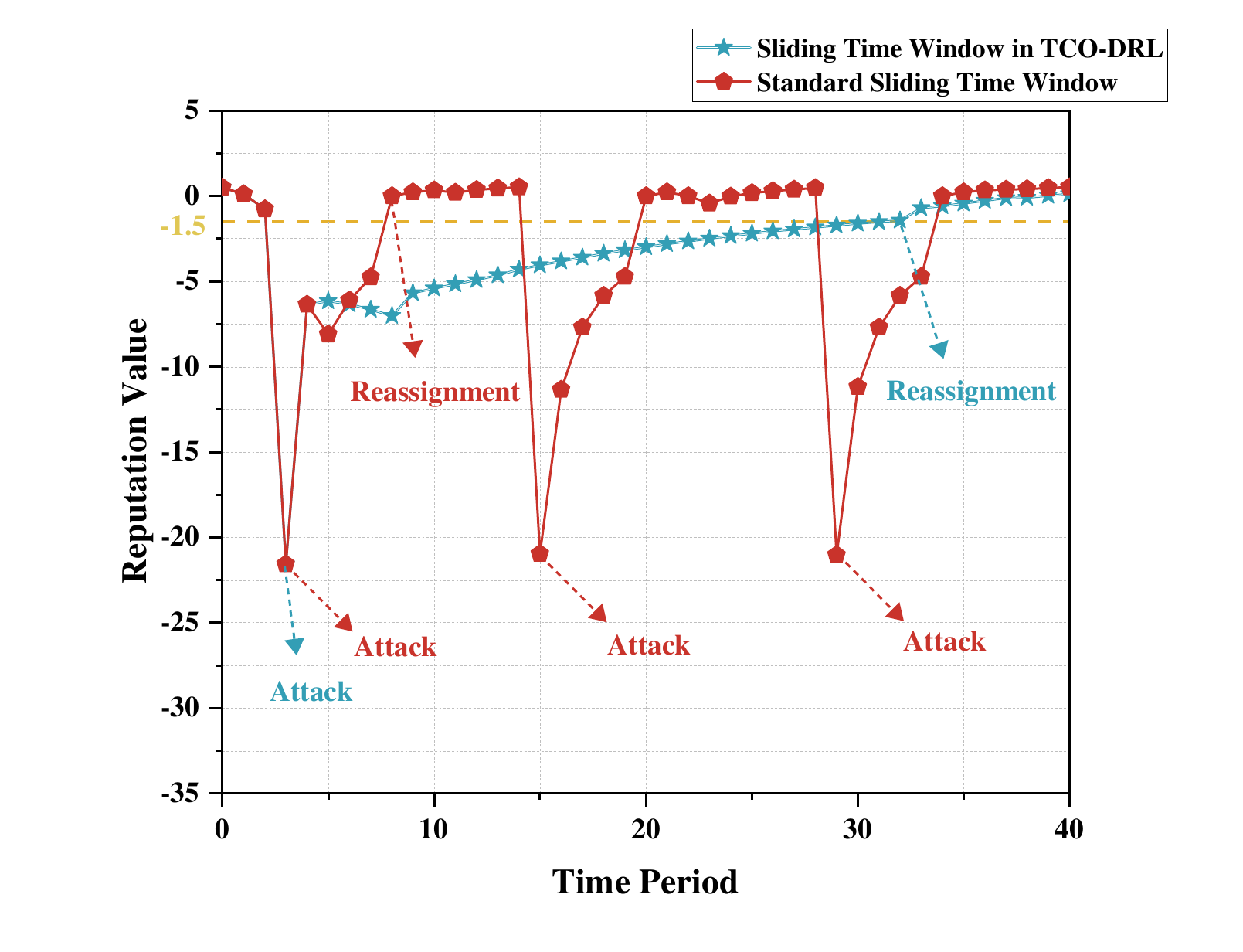}
\caption{Change in reputation value under on-off attack.}
\label{fig_16}
\end{figure}

Next, Fig. \ref{fig_16} demonstrates TCO-DRL's superiority in defending against the on-off attack (OOA). We applied the improved sliding time window to TCO-DRL and compared it with the standard sliding time window. At time period 3, an OOA was initiated. With the standard sliding time window of the same length, the reputation value of the malicious oracle recovered above the trust threshold by time period 8, after just 5 periods, allowing it to continue providing services. In contrast, TCO-DRL makes it difficult for malicious oracles to escape the effects of their harmful actions, making the recovery of trust more challenging. It takes until time period 32 for the reputation to reach above the trust threshold, during which time a malicious node could have committed several malicious actions with the standard sliding time window. Compared to a standard sliding time window of the same length, TCO-DRL extends the reputation recovery period for nodes launching OOAs by 580\%, significantly increasing the time cost of OOAs and effectively reducing their ability to cause harm.

\begin{figure}[!t]
\centering
\includegraphics[width=2.5in]{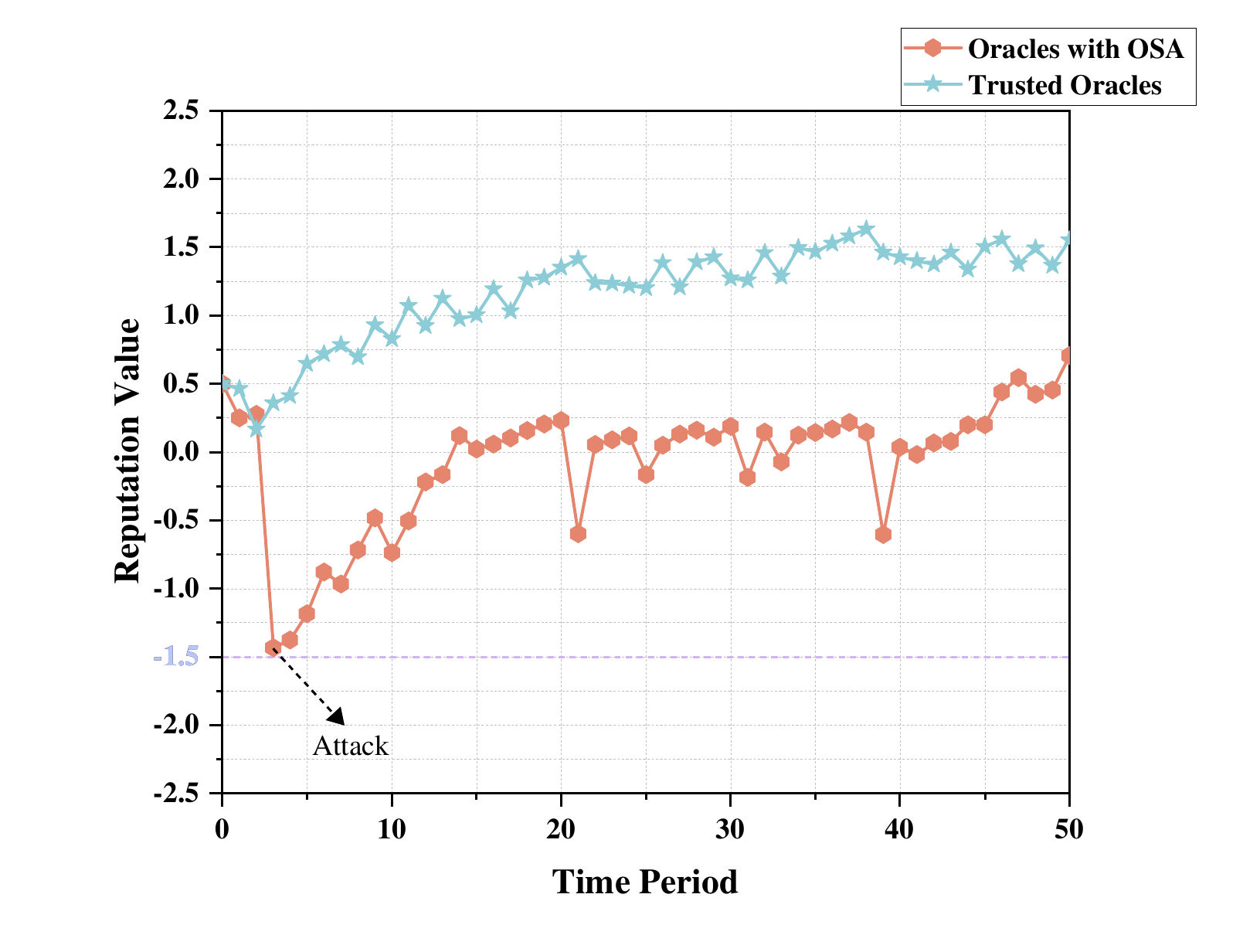}
\caption{Change in reputation value under opportunistic service attack.}
\label{fig_17}
\end{figure}

\begin{figure}[!t]
\centering
\includegraphics[width=2.5in]{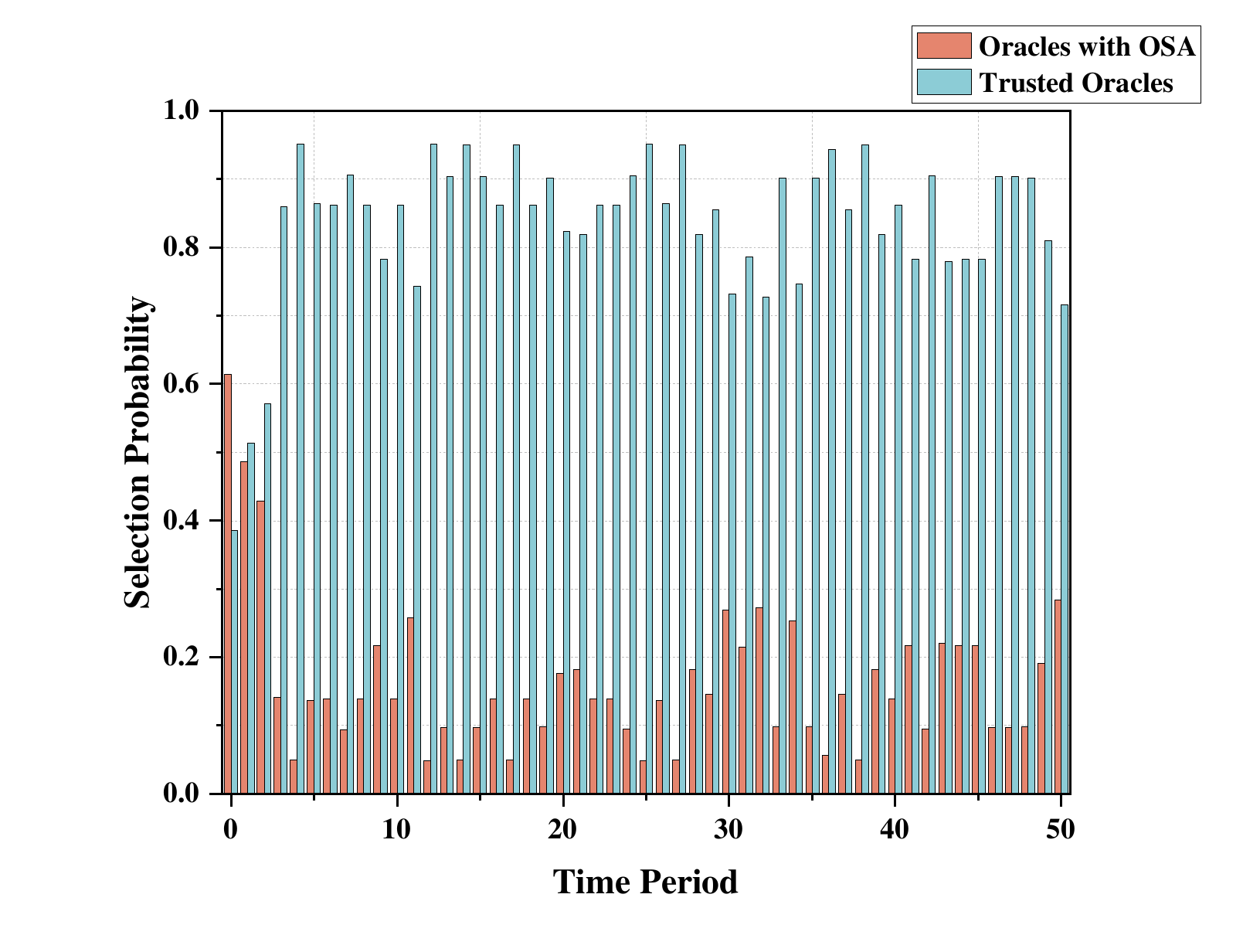}
\caption{Change in selection probability under opportunistic service attack.}
\label{fig_18}
\end{figure}

Finally, we tested TCO-DRL's ability to defend against Opportunistic service attack (OSA). Fig. \ref{fig_17} shows the changes in the reputation values of oracles that could launch OSA compared to those of trusted oracles. In TCO-DRL, once a malicious oracle engages in harmful behavior, its reputation value immediately undergoes a significant decline. Conversely, attempting to quickly boost reputation through short-term positive service is not feasible. In our scheme, maintaining a high reputation requires consistently following the rules and continuously providing high-quality service. 
Although oracles that engage in OSA may still maintain a reputation value above the trust threshold, achieving the reputation level of a trusted oracle requires continuously providing high-quality service. During this process, even minor permissible errors (such as delays caused by network fluctuations) can significantly slow down the growth of their reputation. However, such errors, like network fluctuations, are unavoidable. As shown in Fig. \ref{fig_17}, even after 50 time periods following an OSA, the reputation value of the attacking oracle remains lower than that of a trusted oracle. Fifty time periods is the total number of periods we set, during which 3,000 data requests can be processed. Therefore, if an oracle launches an OSA even once, its reputation will remain lower than that of a trusted oracle throughout the entire service process.
Fig. \ref{fig_18} shows the probability of data request assignments to oracles with OSA and trusted oracles over the corresponding time periods. When an OSA attack is launched at time period 3, there is a significant change in the oracle's reputation value. During the remaining time periods, the reputation values of oracles with OSA are significantly lower than those of trusted oracles, resulting in a very low probability of these oracles being selected to provide services throughout the service process, greatly increasing the difficulty of malicious behavior. This demonstrates TCO-DRL's effectiveness in defending against OSA.

\section{Conclusion}
This paper proposes a blockchain oracle trust-aware and cost-optimized model named TCO-DRL to address the trust issues related to blockchain oracles and the challenge of intelligently selecting oracles in complex, dynamic environments. TCO-DRL provides a comprehensive trust management scheme that evaluates oracle reputation from multiple dimensions. Additionally, TCO-DRL incorporates an improved sliding time window, which not only reduces storage overhead but also significantly weakens the ability of malicious nodes to persistently carry out harmful actions. Based on the DQN algorithm, TCO-DRL effectively identifies malicious oracles and intelligently selects trustworthy and cost-efficient oracles to provide services. Experimental results show that, compared to existing solutions, TCO-DRL significantly reduces the number of data requests allocated to malicious oracles and effectively lowers the average cost. Moreover, TCO-DRL demonstrates strong robustness and resistance against noise and deteriorating environments. By simulating three typical IoT trust-related attacks, TCO-DRL’s effectiveness and strong resilience against attacks were further validated. Future work will focus on introducing incentive mechanisms into the trust management system to encourage oracle nodes to consistently provide reliable services. Additionally, response time will be considered as a key optimization objective to further improve system efficiency and promote multi-objective optimization.

\bibliographystyle{IEEEtran}




%

\begin{IEEEbiography}[{\includegraphics[width=1in,height=1.25in,clip,keepaspectratio]{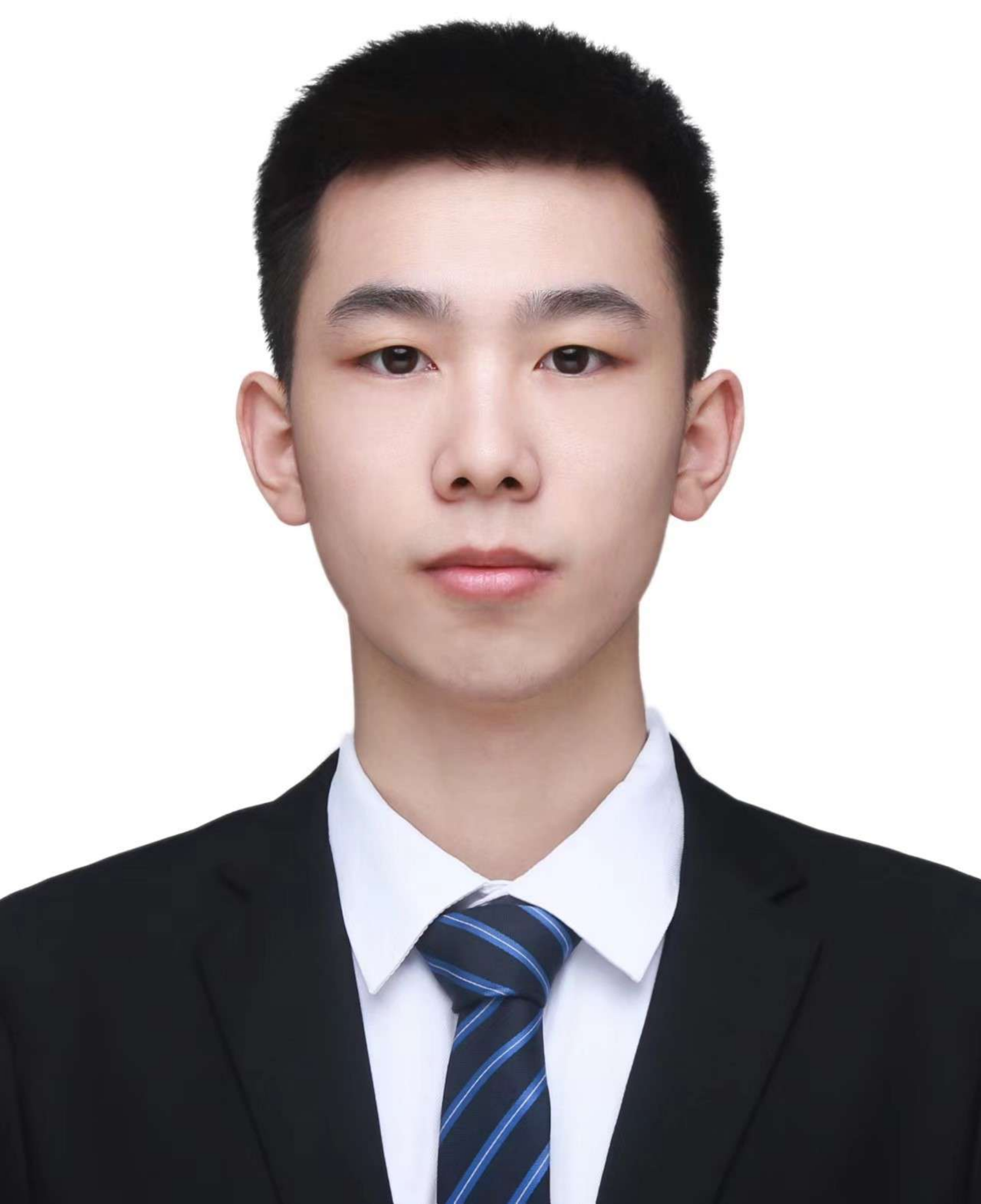}}]{Hengyang Zhang}
Hengyang Zhang received the B.S. degree from North China Electric Power University, Beijing, China, in 2022. He is currently pursuing the M.S. degree with the School of Control and Computer Engineering, North China Electric Power University, Beijing. His research interests include information security, deep reinforcement learning and blockchain systems.
\end{IEEEbiography}

\begin{IEEEbiography}[{\includegraphics[width=1in,height=1.25in,clip,keepaspectratio]{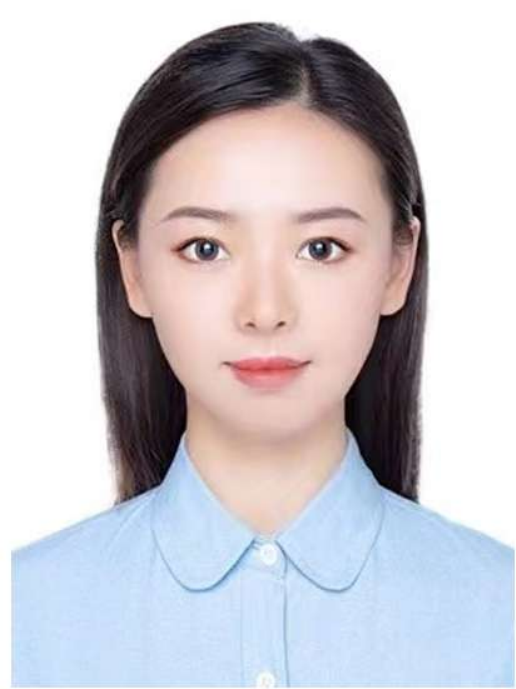}}]{Shike Li}
Shike Li is currently a Lecturer in the School of Computer \& Information Technology at Shanxi University, Taiyuan, China. She received the PhD degree from North China Electric Power University in 2024. She has around 10 papers in IEEE T-ITS, IEEE TVT, IEEE IOTJ, etc. Her research interest includes internet of vehicles, information security, data privacy and blockchain systems. 

\end{IEEEbiography}
\begin{IEEEbiography}[{\includegraphics[width=1in,height=1.25in,clip,keepaspectratio]{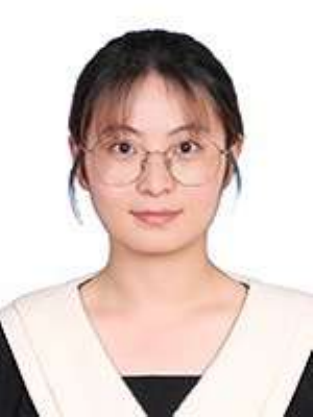}}]{Hang Bao}
Hang Bao is currently a master student in the School of Control and Computer Engineering at North China Electric Power University in Beijing. She received the B.S. degree in Computer Science and Technology from Capital University of Economics and Business in 2023. Her research interests include information security, data privacy and graph neural network.
\end{IEEEbiography}

\begin{IEEEbiography}[{\includegraphics[width=1in,height=1.25in,clip,keepaspectratio]{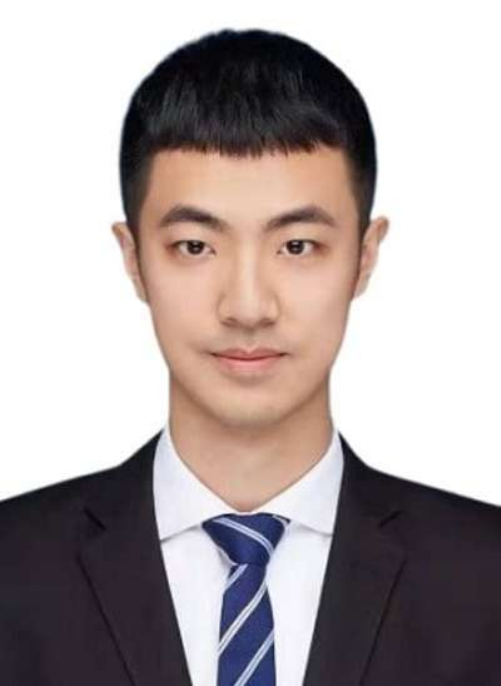}}]{Sixing Wu}
Sixing Wu received the Ph.D. degree in the Department of Electronic Engineering from Tsinghua University in 2021. He focuses on research in the field of natural language understanding, such as sentiment analysis, information extraction and text mining, and has published several papers on journals and conferences in natural language processing and data mining. Besides, his research interests also include the application of artificial intelligence in smart grid and cyber security.
\end{IEEEbiography}

\begin{IEEEbiography}[{\includegraphics[width=1in,height=1.25in,clip,keepaspectratio]{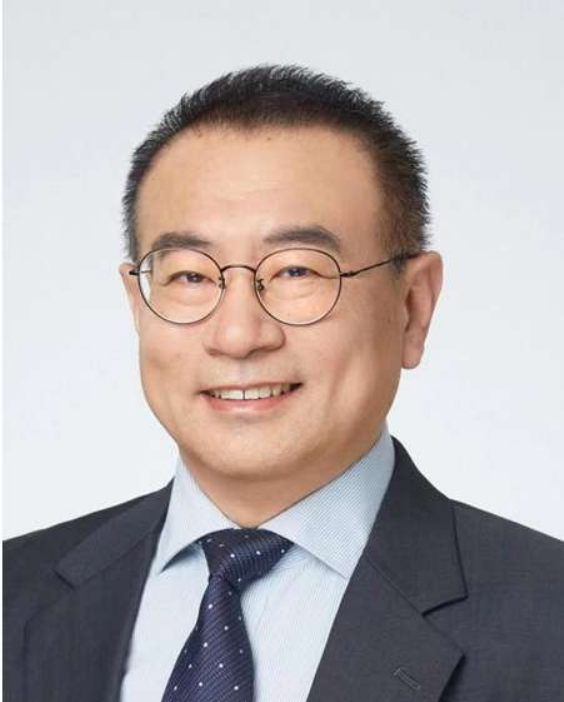}}]{Jianbin Li}
Jianbin Li received the BS degree from Tsinghua University in 1992, and the MS degree from Analysis and Forecast Center of the State Seismological Bureau in 1995. He was the dean of the Institute of Information Security and Big Data at Central South University. He is currently a Full Professor with the North China Electric Power University in Beijing. His research interests mainly include information security, big data systems and blockchain.
\end{IEEEbiography}








\end{document}